\documentclass[preprintnumbers,amsmath,amssymb,floatfix,twocolumn,showpacs,prb]{revtex4-1}

\usepackage{dcolumn}
\usepackage{bm}
\usepackage{graphicx}
\usepackage{float}
\usepackage{hyperref}

\begin{document}


\title{Geometry of Landau orbits in the absence of rotational symmetry}

\author{F. D. M.  Haldane and Yu Shen}
\affiliation{Department of Physics, Princeton University,
Princeton NJ 08544-0708}

\date{May 15, 2016}
\begin{abstract}
The integer quantum Hall effect (IQHE) is usually modeled using  Galilean-invariant
or rotationally-invariant Landau levels.   However, these are not generic symmetries
of electrons moving in a crystalline background, even in the low-density continuum
limit. We present a treatment of the IQHE which abandons Galilean
dispersion and rotational symmetry, only keeping inversion symmetry,
which exposes topological properties and more generic geometric structures.
 We define an emergent metric $g^n_{ab}$ for each Landau level with a
 reformulation of the Hall viscosity. The metric is then used to
 define guiding-center coherent states where the corresponding single-particle wavefunctions are holomorphic functions of $z^*$ times a Gaussian. By numerically studying cases with quartic dispersion, we show that the number of the zeroes
of the wavefunction encircled by the semiclassical orbit, denoted by
$n$, defines a  ``topological spin'' $s_n$ by $s_n=n+\frac{1}{2}$,
while its original definition as an ``intrinsic angular momentum'' is no
longer valid without rotational symmetry. We also  present results for
the density and current responses which differentiate between diagonal
and Landau-level-mixing terms. 
\end{abstract}
\maketitle

\section{Introduction}
Most treatments of the integer and fractional  quantum Hall effects
(IQHE and FQHE)  in
the low-density two-dimensional electron gas (2DEG) subjected to a
uniform magnetic flux density begin by making  the
simplifying assumption that the electrons obey Newtonian dynamics with
2D Galilean invariance, and that the system has  a continuous rotational
symmetry around an axis  normal to the 2D surface on which the electrons
move.   

Among the simplifications that follow from this is the
well-known result that when rotational symmetry is present, the
 wavefunction of an electron in the lowest
Landau level is a holomorphic function $f(z)$ = $f(x + iy)$ times a
Gaussian factor, which played a key role in Laughlin's discovery\cite{laughlin} of a
model wavefunction for the FQHE with Landau-level filling $\nu$
=  $1/m$.     The apparent importance of rotational symmetry was
further reinforced when it was observed\cite{Moore} that the Laughlin state wavefunctions
were mathematically related to ``conformal blocks'' of (Euclidean) 2D conformal
field theory in which the global rotation symmetry $SO(2)$  is promoted to a
local Euclidean conformal symmetry $CO(2)$.      

Much of the interest in the
QHE relates to its robust  topological properties, which by definition are not
related to the presence or absence of symmetries such as rotational,
Galilean or conformal symmetries,  so the presence of such symmetries as a simplifying ``toy
model'' feature is not problematic. 
However  there has been recent interest
in geometrical properties of the QHE\cite{Bradlyn1, Bradlyn2, Abanov,Gromov1,Gromov2,Can,Cho,Gromov3},  in which symmetries play an
important role, and the inclusion of non-generic symmetries may in this
case be 
misleading.

The assumption of a local Galilean structure with an  inertial
mass tensor  given by a scalar gravitational mass times the
Euclidean metric of the flat plane has been extended by some authors
to a  treatment of the QHE that places the electrons  on a Riemann surface
that can be thought  of as a curved 2D surface immersed in 3D
Euclidean space, from which it receives an induced metric that defines
a Laplacian operator.  A related
formulation of the problem has been given as ``Newton-Cartan'' theory\cite{Son}.
However the QHE is experimentally exhibited by electrons moving on a
lattice plane inside crystalline condensed matter, in which there is
generically no Galilean invariance or rotational symmetry, and the 
non-Newtonian
electron
kinetic energy derives from the  band structure of an underlying
atomic-scale crystal structure.       

Models of the QHE that include
non-generic symmetries are certainly useful  models which will
exhibit generic topological behavior, but cannot be relied upon to
exhibit generic geometrical behavior.      A useful analogy is the
theory of metals, for which the simple model of a Galilean-invariant 
Fermi gas with a spherical Fermi surface is often used as an
initial ``toy model'' that exhibits many generic features of metals
(such as  a $T$-linear specific heat),  but fails to capture the intricate geometry
of true metallic Fermi surfaces.

In an attempt to clarify the nature of geometric properties of the QHE,
such as  the ``Hall viscosity''\cite{ASZ},  we will here present a
treatment of the integer QHE which abandons the Galilean 
dispersion relation of non-relativistic Newtonian dynamics,
where
\begin{equation}
\varepsilon (\bm p) = \frac{p_x^2 + p_y^2}{2m}
\label{gal}
\end{equation}
keeping only the weaker condition
\begin{equation}
\varepsilon (-\bm p) = \varepsilon (\bm p)
\label{invert1}
\end{equation}
and the simplification that the classical Landau orbits  $\varepsilon
(\bm p)$ = $E$, constant, are compact closed curves in 2D momentum
space, so that classical Landau orbits are uniquely fixed by their energy and
guiding center.   The Galilean-invariant model is a special case of
these weaker  restrictions, which  simplify the semiclassical
quantization.
The weaker restrictions we impose  could
themselves be dropped at the expense of introducing some additional
complications, but we believe that as they permit the complete removal
of the non-generic Galilean and rotational-symmetry structures from
the theory of the QHE,  they allow more  generic geometrical structures
to be revealed.

The ``momentum space inversion symmetry'' (\ref{invert1}) can be
justified as a condition that derives from the assumption that the
2DEG has unbroken time-reversal symmetry in the absence of an applied
magnetic field.    It is also compatible with spatial inversion
symmetry, which is the only metric-independent point symmetry compatible with generic
crystalline condensed matter, as it is the only point symmetry of a
generic Bravais lattice.

In this work, the geometrical description will be restricted to that of the integer QHE of
non-interacting electrons in non-Galilean Landau levels, and the
extension to the richer structure of the FQHE will be given elsewhere,
building on the IQHE ingredients described here.

This paper is organized as follows. In Section \ref{sec:notation} we briefly introduce a formulation of
2D spatial geometry using consistent covariant and
contravariant index placement. With the algebra of Landau orbits and
guiding centers reviewed in Section \ref{sec:algebra}, we study the
geometry of Landau orbits by emphasizing on a unimodular metric
$\tilde g_{ab}$ in Section \ref{sec:geometry}. Then in
Section\ref{sec:viscosity} we provide a natural choice of the metric
along with a reformulation of the Hall viscosity in the absence of
rotational symmetry. In Section \ref{sec:zeroes} we define a topological
``Landau orbit spin'' $s_n$ by investigating the root structure of the
wavefunction,  and the case of a  quartic dispersion is numerically
studied in Section \ref{sec:casestudy}. After a brief discussion of some
other properties of the Landau orbits in Section \ref{sec:properties}, we
describe in Section \ref{sec:response} generic results for the density and current responses in the absence of rotational symmetry. 

\section{2D geometry with consistent index placement}
\label{sec:notation}
The Euclidean plane can be
parametrized by a
Cartesian coordinate system 
\begin{equation}
\bm x = x^a\bm e_a, \quad \bm e_a \cdot \bm e_b = \delta_{ab},
\end{equation}
where $\{\bm e_a, a=1,2\}$ is a basis of $\bm x$-independent orthonormal tangent vectors, and
\begin{equation}
\bm e_a \times \bm e_b = \epsilon_{ab}\bm n,
\end{equation}
where $\bm n$ is the unit normal to the plane, which is also $\bm x$-independent.
Here $\delta_{ab}$ is the 2D Euclidean metric induced on the plane by
its embedding in 3D Euclidean space, and $\epsilon_{ab}$ is the
antisymmetric  2D Levi-Civita symbol, which has a sign ambiguity
because the unit normal is only defined up to $\pm \bm n$ (the
\textit{orientation} choice).
 
It will prove very useful to use a formalism that distinguishes
\textit{vectors}, such as velocities $\bm v$ = $v^a\bm e_a$ =  $\dot x^a\bm
e_a$ which have components with upper indices (contravariant to those
of $\bm e_a$), from \textit{covectors}, such as forces,  which have lower
(covariant) indices $F_a$ $\equiv$  $\bm e_a \cdot \bm F$.    Then the
scalar product $\bm F\cdot \bm v$ $\equiv$ 
 $F_av^a$ $\equiv$ $F_1v^1 + F_2v^2$  is only defined as a
bilinear relation between vectors and covectors, with Einstein
summation convention on repeated upper/lower index pairs.
In this formalism, the Euclidean metric $\delta_{ab}$, the inverse
metric $\delta^{ab}$ and the Kronecker symbol $\delta^a_b$ are
distinguished, although they are numerically equal when  a Cartesian
coordinate system is used.   

The dual Levi-Civita symbol
$\epsilon^{ab}$ is also numerically-equal to (but distinguished from)
$\epsilon_{ab}$.  Technically, $\epsilon_{ab}$ is a covariant
pseudotensor density with weight $-1$,
and  $\epsilon^{ab}$ is a contravariant  pseudotensor
density with weight $+1$,   
If  $g_{ab}$ is a metric tensor,
\begin{equation}
g \equiv \det g = \epsilon^{ac}\epsilon^{bd}g_{ab}g_{cd}
\end{equation}
is a scalar density with weight 2.   A useful formal identity is
\begin{equation}
\epsilon^{ab}\epsilon_{cd} = \delta^a_c\delta^b_d -
\delta^a_d\delta^b_c.
\end{equation}
In any expression, covariant indices, contravariant indices, and
tensor weights must balance.

The use of consistent index placement may seem pedantic
in systems where the Euclidean metric plays a fundamental role, such
as those governed by Newtonian dynamics, in which the inertial mass tensor of a particle $m_{ab}$ is
given by $m\delta_{ab}$, where $m$ is its gravitational mass,
However, in \textit{generic} non-relativistic crystalline condensed matter
(\textit{i.e.}, with a trigonal as opposed to 
a cubic lattice, and no point-group symmetry other than spatial
inversion)
the Euclidean metric \textit{plays no role whatsoever in the
  non-relativistic physics},
except for the definition of the non-relativistic limit
$\delta_{ab}v^av^b \ll c^2$, where $\bm v$ = $v^a\bm e_a$ is the
velocity of an electron relative to the preferred inertial frame of
condensed matter, in which the underlying atomic lattice  is at rest.

Consistent index placement exposes the presence of the
Euclidean metric in physical equations: for example, the azimuthal angular
momentum, which is conserved in Newtonian dynamics on the plane, is given by
\begin{equation}
L = \epsilon_{ab}\delta^{bc}x^ap_c.
\end{equation}
The presence of the inverse Euclidean metric (required to balance the
index placement) immediately identifies the angular momentum
as a quantity  without validity in a generic crystalline environment.

\section{The algebra of Landau orbits and guiding centers}
\label{sec:algebra}
The Schr\"odinger representation of the momentum of an electron is
\begin{equation}
\bm e_a \cdot \bm p = p_a = -i\hbar \frac{\partial}{\partial x^a} -
eA_a(\bm x),
\end{equation}
where 
\begin{equation}
\frac {\partial  A_b}{\partial x^a} 
-
\frac {\partial  A_a}{\partial x^b} 
= \epsilon_{ab} B(x), 
\end{equation}
and $x^a$ are the coordinates of the electron.
With this definition, the Levi-Civita symbol is odd under
time-reversal.
We will concentrate on the case $B(\bm x)$ = $B$, constant, leaving
open the eventual extension to include a slow spatial variation $B(\bm x)$.
In the case of constant $B$,  the commutation relations of the components of the
momentum have the simple form of a Heisenberg algebra
\begin{equation}
[p_a,p_b] = i\hbar eB\epsilon_{ab},
\label{pp}
\end{equation}
which has 
 the exponentiated form
\begin{equation}
U(\bm a_1)U(\bm a_2)\ldots U(\bm a_n) =
\prod_{i<j} e^{\frac{1}{2}i\varphi(\bm a_i, \bm a_j)}U({\textstyle\sum_i} \bm a_i),
\end{equation}
 where $U(\bm a)$ = $\exp( i \bm a \cdot \bm p)$  with real $a^a$
are unitary operators, and 
\begin{equation}
\varphi (\bm a,\bm a') = -\hbar eB \epsilon_{ab}a^aa'^b.
\end{equation}

The remaining commutation relations are unaffected by $B$:
\begin{equation}
[x^a,x^b] = 0, \quad [x^a,p_b] = i\hbar \delta^a_b.
\end{equation}
In a system with uniform magnetic flux density, the electron coordinate may be
decomposed as
\begin{equation}
\bm x = \bm R + (eB)^{-1}\epsilon^{ab}\bm e_a p_b,
\end{equation}
 where
$\bm R$ = $R^a\bm e_a$ is the \textit{guiding center} of the Landau
orbit,
with commutation relations
\begin{equation}
[R^a,R^b] = -i\hbar(eB)^{-1}\epsilon^{ab}, \quad [R^a,p_b]  = 0,
\end{equation}
and which defines a second independent Heisenberg algebra. 
It is useful to write
\begin{equation}
p_a = B\epsilon_{ab}d^b,
\end{equation}
where
\begin{equation}
\bm d = e(\bm x - \bm R)
\end{equation}
can be interpreted as the electric dipole moment  defined by  the displacement of the
electron relative to its  guiding center.   The inversion
symmetry $\varepsilon(\bm p)$ = $\varepsilon(-\bm p)$ means that in a
Landau orbit eigenstate $|\psi_{n\alpha}\rangle$ of $\varepsilon(\bm p)$, the expectation value
of this electric dipole vanishes, and 
\begin{equation}
\langle \psi_{n\alpha}|\bm p|\psi_{n\alpha}\rangle = 0,
\end{equation}
which provides an unambiguous definition of the guiding center independent of the Landau level.    

In a uniform magnetic field, a representation of the algebras of $p_a$
and $R^a$ can be  given in terms of \textit{independent harmonic oscillator degrees
of freedom}
\begin{eqnarray}
p_a &=& \surd ({\textstyle\frac{1}{2}}\hbar |eB|)\left ( \tilde e_a a^{\dagger} + \tilde
        e_a^*a\right ) ,\nonumber \\
R^a &=& \surd ({\textstyle\frac{1}{2}}\hbar |eB|^{-1}) \left (\bar e^a b + \bar
        e^{a*}b^{\dagger}\right ), 
\label{harmonic}
\end{eqnarray}
where
\begin{equation}
\epsilon^{ab}\tilde e_a^*\tilde e_b = \epsilon_{ab}\bar e^{a*}\bar e^b = i, 
\label{uni}
\end{equation}
and 
$[a,a^{\dagger}]$ = $[b,b^{\dagger}]$ = 1, $[a,b]$ =$[a,b^{\dagger}]$
= 0. 
A metric is said to be \textit{unimodular} if it has determinant 1; 
The condition (\ref{uni}) means that \textit{two} unimodular
Euclidean-signature metrics  and their inverses are defined by
\begin{eqnarray}
\tilde e^*_a\tilde e_a &=& {\textstyle\frac{1}{2}}\left ( \tilde
                           g_{ab} + i \epsilon_{ab}\right ), \quad
\tilde g^{ab} = \epsilon^{ac}\epsilon^{bd}\tilde g_{cd},  \nonumber\\
\bar e^{a*}\bar e^a &=& {\textstyle\frac{1}{2}}\left ( \bar
                           g^{ab} + i \epsilon^{ab}\right ), \quad 
\bar g_{ab} = \epsilon_{ac}\epsilon_{bd}\bar g^{cd} .
\end{eqnarray}
The standard treatments would immediately  set both $\tilde g_{ab}$ and $\bar
g_{ab}$ equal to  the Euclidean metric $\delta_{ab}$, but we emphasize that
they are independent free parameters of  the representation (\ref{harmonic}).

It is now useful to introduce the  generators of linear
area-preserving $SL(2,R)$ linear deformations of momentum space
\begin{equation}
p_a \mapsto  U(\beta)p_aU(\beta)^{-1} =\Lambda_a^{\; b} (\beta)p_b, 
\label{unitary}
\end{equation}
where
\begin{equation}
\det \Lambda
\equiv 
{\textstyle\frac{1}{2}}\epsilon^{ac}\epsilon_{bd}\Lambda_a^{\;b}
\Lambda_c^{\;d} = 1.
\end{equation}
This unitary transformation leaves the guiding centers invariant, and preserves the commutation relation
(\ref{pp}).  The unitary operator $U(\beta)$ is given by
\begin{equation}
U(\beta ) = \exp ( i \pi  \beta^{ab}\gamma_{ab}), 
\end{equation}
where $\beta_{ab}$ = $\beta_{ba}$ is real, and the 
Hermitian  generators $\gamma_{ab}$ = $\gamma_{ba}$ are given by
\begin{equation}
{\textstyle\frac{1}{4}}\{p_a,p_b \} =  \hbar eB \gamma_{ab} .
\label{rep}
\end{equation}
They obey the  non-compact Lie algebra
\begin{equation}
[\gamma_{ab},\gamma_{cd}] = {\textstyle\frac{1}{2}}i\left (
\epsilon_{ac}\gamma_{bd} +
\epsilon_{bc}\gamma_{ad} +
\epsilon_{ad}\gamma_{bc} +
\epsilon_{bd}\gamma_{ac} \right )
\label{Lie}
\end{equation}
with a quadratic Casimir
\begin{equation}
C_2 = \det \gamma =  {\textstyle\frac{1}{2}}
\epsilon^{ac}\epsilon^{bd}\gamma_{ab}\gamma_{cd}.
\end{equation}
The representation (\ref{rep}) of this algebra has
$C_2$ = $-\frac{3}{16}$.
Any unimodular metric $\tilde g_{ab}$ with $\det \tilde g$ = 1 defines a Cartan subalgebra
generated by $L_g$ = $\tilde g^{ab}\gamma_{ab}$.
The explicit form of this  representation
in terms of harmonic oscillator  operators  (\ref{harmonic})
is
\begin{eqnarray}
\gamma_{11} &=& {\textstyle\frac{1}{4}} \left (a + a^{\dagger}\right)^2\nonumber\\
\gamma_{22} &=& -{\textstyle\frac{1}{4}} \left (a - a^{\dagger}\right)^2\nonumber\\
\gamma_{12} &=& {\textstyle\frac{1}{4}} i\left  ((a)^2 -
                (a^{\dagger})^2\right ).
\end{eqnarray}
Then one choice of a Cartan subalgebra is
\begin{equation}
L = \gamma_{11} + \gamma_{22} =  {\textstyle\frac{1}{2}}(a^{\dagger}a +
                              aa^{\dagger}).
\end{equation}
This representation is in fact reducible into two irreducible
representations of the Lie algebra (\ref{Lie}) with the same value of the quadratic  Casimir, distinguished
by parity under $a \mapsto -a$.

The transformation   (\ref {unitary}) is a  hyperbolic ``squeeze
mapping''
of momentum space
if $\det \beta < 0$, a parabolic shear if $\det \beta $ = 0, and an
elliptic  ``generalized
rotation'' through an angle $\theta $ = $\pi \surd (\det\beta)$ if $\det
\beta > 0$.   For all $\beta_{ab}$ with  $\det \beta  $ =
1,  $U(\beta)$ has the property $U(\beta)^2$ = $-1$, and reduces to the  
inversion mapping 
\begin{equation}
U(\beta) p_a U(\beta) = - p_a,  \quad \det \beta = 1.
\end{equation}
This leaves $\gamma_{ab}$
unchanged, and  thus, like the quadratic Casimir,  $U(\beta)$ with
$\det \beta $ = $1$
is in the center of the Lie algebra.
The full $SL(2,R)$ group is thus spanned by  $U(\beta)$ with
\begin{equation}
\det \beta \le 1,
\end{equation}
where all transformations with $\det \beta$ = 1 are equivalent,
and equal to the inversion.

The analogous algebra that acts on the guiding centers has generators
$\bar  \gamma_{ab}$ defined by
 \begin{equation}
{\textstyle\frac{1}{4}} \{R^a,R^b\} =  -\hbar (eB)^{-1} \epsilon^{ac}\epsilon^{bd}\bar \gamma_{cd}.
\end{equation}
This obeys the algebra (\ref{Lie}) with the same quadratic Casimir.

\section{Landau orbits and their geometry}
\label{sec:geometry}

In keeping with the interpretation that  electronic  kinetic energy $\varepsilon (\bm p)$ derives
from an expansion of a band-structure  around a band minimum, 
it will be assumed that
\begin{equation}
\varepsilon(\bm p)  = \varepsilon(-\bm p)  > \varepsilon(\bm 0) ,\quad \bm p \ne \bm 0.
\end{equation}
A  ``\textit{regularity condition}'' on $\varepsilon(\bm p)$
will also be assumed:  this is that  $\varepsilon(\lambda \bm p)$ is
analytic in $\lambda$ for all $\bm p \ne \bm 0$.
In fact, for an operator $\varepsilon (\bm p)$ to remain well-defined
when it is a bivariate function of two  non-commuting components $p_a$
that obey a Heisenberg algebra, 
it appears that $\varepsilon(\bm p)$ should not just be
analytic in each component of $\bm p$, 
but should be an \textit{entire} function, so that, for
any  $c$-number  $\bm p_0$, 
$\varepsilon(\bm p_0 + (\bm p - \bm p_0))$ has an
absolutely-convergent expansion in 
powers of the components
of  $\delta\bm  p\bm (p_0)$ =  $\bm p-\bm p_0$,
of the form
\begin{equation}
\varepsilon(\bm p) =
\sum_{n=0}^{\infty} 
 \varepsilon_n^{a_1a_2\ldots a_n}(\bm p_0) \{
\delta p_{a_1}(\bm p_0) ,\ldots, \delta p_{a_n}(\bm p_0)\}
\end{equation}
(where $\{\ldots \}$ is a symmetrized product of the non-commuting variables)
which is independent of  the (arbitrary) choice of the expansion 
origin $\bm p_0$.
This ``strong regularity'' condition is satisfied
if $\varepsilon(\bm p)$ is a polynomial in the components $p_a$,
which is the case investigated here.

A   ``\textit{strict monotonicity  condition}'' will also  be  assumed
so that,
for all $\bm p \ne \bm 0$,  
$\varepsilon (\lambda \bm p) $ is a strictly  monotonically-increasing
function in the range $\varepsilon(\bm 0) \le
\varepsilon(\lambda \bm p) <\infty$  for $0\le \lambda < \infty$,
that is unbounded as $\lambda \rightarrow \infty$.
This implies that all  curves $\varepsilon (\bm p)$ = $E >
\varepsilon(\bm 0)$ are compact closed
curves in the momentum plane $(p_1,p_2)$ $\in \mathcal R_2$,  with the topology of a
circle
$S_1$, and enclose the origin.  These can be parametrized as $\bm p(E,
\theta)$  = $\bm p(E, \theta + 2\pi)$  and strict monotonicity implies
that the sense of the orbit can be chosen so that, for all $\theta$,
\begin{equation}
w ( E, \theta) \equiv \epsilon^{ab}p_a\partial_{\theta}p_b(E, \theta) > 0,
\end{equation}
where $\partial_{\theta} \bm p(E, \theta)$ $\equiv$ $\partial \bm
p(E, \theta)/\partial \theta$.

The group velocity of the dispersion is
\begin{equation}
\bm v (\bm p) = \bm e_a\frac{\partial \varepsilon}{\partial p_a},
\end{equation}
where $\bm v(-\bm p)$ = $-\bm v(\bm p)$.
The strict monotonicity and regularity conditions imply that the 
the only solution of the condition  $\bm v(\bm p)$ = $\bm 0$
is $\bm p$ = 0, and that $\bm p \cdot \bm v(\bm p)$ $\equiv$ $p_av^a(\bm p) >0$ for $\bm p \ne  0$.
The classical Hamiltonian equation of motion is
\begin{equation}
\frac{d\bm x} {dt} = \bm v(\bm p), \quad 
\frac{dp_a}{dt} = eB \epsilon_{ab}v^b(\bm p)
\end{equation}
and a classical particle with kinetic energy $E$ moves on the closed
momentum-space trajectory
$\bm p(E,\theta)$ with
\begin{equation}
\frac{d\theta}{dt} = 
\frac{eBp_av^a(\bm p)}
{w(E,\theta)} > 0.
\end{equation}
It is worth stressing that,  in marked contrast to the case of
Newtonian dynamics in free space, where $\varepsilon (\bm p)$ = 
$\frac{1}{2}m^{-1}\delta^{ab}p_ap_b$, the
Euclidean metric $\delta_{ab}$ has no place in the
Hamiltonian equations of motion of  non-relativistic Landau orbits of
electrons inside condensed matter.

While the classical Landau-orbit dynamics is fundamentally
Hamiltonian, a classical Lagrangian
description becomes possible with an additional restriction on the 
function $\varepsilon(\bm p)$, that the odd function $\bm v(\bm p)$
defines an invertible map from the full momentum plane  to a compact
region $\mathcal D_v$ of the velocity plane, so that the  kinetic energy function
has the form $\varepsilon(\bm p)$ = 
$\tilde \varepsilon (\bm v(\bm p))$, where 
$\tilde \varepsilon(\dot{ \bm x})$, $\dot{\bm x} \in \mathcal D_v$, is the Lagrangian kinetic energy.

The momentum-space area $\mathcal
A(E)$
 enclosed by the curve  $\bm p(E, \theta)$ is
\begin{equation}
\mathcal A(E) ={\textstyle\frac{1}{2}}
\int_0^{2\pi} d\theta \epsilon^{ab}p_b\partial_{\theta} p_b(E, \theta).
\end{equation}
The strict monotonicity condition means that $\mathcal A(E)$ is a 
strictly monotonically increasing unbounded function of $E \ge
\varepsilon(\bm 0)$, with
$\mathcal A(E) \rightarrow 0$ as $E\rightarrow \varepsilon(\bm 0)^+$.
The semiclassical quantization of Landau levels is
\begin{equation}
\mathcal A(E_n)  = 2\pi\hbar eB  s_n > 0, \quad (-1)^{2s_n}  = -1.
\label{semiclassical}
\end{equation}
The monotonicity condition
ensures that  for 
\begin{equation}
s_n = (n +
{\textstyle\frac{1}{2}}){\rm sgn}(eB), \quad n = 0,1,2,\ldots,
\end{equation}
$E_{n+1} > E_n$,  and    $n$
coincides with the conventional Landau-level index.

Because the components of $\bm p$ no longer commute with each other
when $\hbar eB \ne  0$,
the one-particle quantum Hamiltonian must be given as a series expansion in $p_a$
with an ordering convention.    To reflect our choice to specialize to
inversion-symmetric models, which singles out $\bm p_0$ = 0 as a
preferred origin for an expansion in $\bm p - \bm p_0$,
 the form used here will be
\begin{equation}
H =  \sum_{n=0}^{\infty} \frac{1}{2n!}  (p_{a_1}\ldots
p_{a_{n}} )A_n^{\{a_i\},\{b_i\}} (p_{b_1}\ldots p_{b_n}),
\label{ham}
\end{equation}
where $A_n^{\{a_i\},\{b_i\}}$ is symmetric under any permutations
of indices within the sets $\{a_1,\ldots a_n\}$ or $\{b_1,\ldots ,b_n\}$.
In the limit $\hbar eB \rightarrow 0$, $H$ becomes the function
$\varepsilon(\bm p)$ in the momentum plane, which is then given in
terms of the expansion coefficients $A_n^{\{a_i\},\{b_i\}}$.   This
form remains Hermitian if translational invariance is broken to make
$A_n^{\{a_i\},\{b_i\}}(\bm x)$ or $B(\bm x)$  adiabatically-varying functions of position.

As a result of translational invariance, the Landau levels are macroscopically degenerate, with one independent
state for each London quantum $\Phi_0$ = $\hbar /e$ of magnetic flux
passing through the plane:
\begin{equation}
H|\Psi_{n\alpha}\rangle = E_n(\hbar eB)|\Psi_{n\alpha}\rangle.
\end{equation}
It will be useful to generalize this Hamiltonian eigenvalue problem to
a \textit{Lagrangian} eigenvalue problem
\begin{eqnarray}
\hat L(\bm v) &\equiv& \bm v \cdot \bm p - H, \nonumber \\
\hat L(\bm v) |\Psi_{n\alpha}(\bm v) \rangle
&=& L_n(\bm v,\hbar eB)
|\Psi_{n\alpha}(\bm v) \rangle,
\end{eqnarray}
where the components of the velocity $\bm v$ are  $c$-number parameters
of the Hermitian operator $\hat L(\bm v, \hbar eB)$, and the eigenvalue $L_n(\bm
v)$ is a function of $\bm v$ that characterizes each Landau level, and
will be the Lagrangian for the motion of the guiding center of an
electron in that level.
For the Galilean-invariant Newtonian problem, this Lagrangian reduces to the
simple form
\begin{equation}
L_n(\bm v, \hbar eB) =  {\textstyle\frac{1}{2}}m\delta_{ab}v^av^b - m^{-1}\hbar
eB s_n.
\end{equation}

The eigenstates of the Hamiltonian can be represented  in a basis of
harmonic oscillator states
\begin{eqnarray}
|\Psi_{n\alpha}\rangle &=& \sum_{m=0}^{\infty} u_{n,m} \frac{(a^{\dagger} )^m}{\surd
  (m!)} |\phi_{\alpha}\rangle \equiv   f_n(a^{\dagger})
|\phi_{\alpha}\rangle   , \nonumber \\
a|\phi_{\alpha}\rangle &=& 0,
\end{eqnarray}
where both $f_n(a^{\dagger})$ and the reference state $|\phi_{\alpha}\rangle $ depend on the
choice of unimodular metric $\tilde g_{ab}$.    Note that while the
diagonalization is carried out in a basis defined by a given choice
of this metric, it will turn out to be useful to represent each 
Landau-level eigenstate in this form using a Landau-level-specific
choice  $\tilde g_{ab} \propto\tilde g^n_{ab}$, where $g^n_{ab}$ is a
(non-unimodular) metric specific to the Landau level.

The question now arises, what is the natural basis to use to resolve
the degeneracy of the Landau levels?
Since
\begin{equation}
[H, \bm R] = 0,
\end{equation}
the guiding center is a classical constant of the motion.  However,
quantum mechanically, its components do not commute, and obey an
uncertainty principle
\begin{equation}
\bar g_{ab}\left ( \langle \Psi| (R^aR^b)|\Psi\rangle -
\langle \Psi|R^a|\Psi\rangle \langle \Psi|R^b|\Psi\rangle \right )
\ge \ell_B^2,
\label{coh}
\end{equation}
where $\bar g_{ab}$ is any Euclidean-signature unimodular metric,
and
\begin{equation}
\ell_B = \surd \left (\hbar |eB|^{-1}\right ).
\end{equation}

The closest
quantum mechanical state to the classical state with a fixed guiding
center
is a  \textit{guiding-center coherent state}  $|\Psi_n(
\bar{\bm  x},\bar g)\rangle $ that obeys the inequality  (\ref{coh}) as an
\textit{equality}, with
\begin{equation}
\langle \Psi_n(\bar {\bm x}, \bar g)|\bm R| \Psi_n(\bar {\bm x},
\bar g)\rangle = \bar {\bm x}.
\end{equation}
The standard  treatments of Landau levels based on a Newtonian kinetic
energy  $\frac{1}{2}m\delta_{ab}v^av^b$ generally use a coherent state
with $\tilde g_{ab}$ given by the Euclidean  metric $\delta_{ab}$, but since non-Newtonian
Hamiltonian dynamics has no dependence on the Euclidean metric, this
is no longer the ``obvious'' choice.    Below, we will show that there
is a ``natural'' choice that in general is different for each Landau
level, but since \textit{any} unimodular metric is a viable choice, we
leave it unspecified for now, and use it to set both $\tilde g_{ab}$
and $\bar g_{ab}$, so $\bar g_{ab}$ = $\tilde g_{ab}$.

A unimodular metric defines a \textit{complex structure}  through the
factorization
\begin{equation}
{\textstyle\frac{1}{2}}\left ( \bar g_{ab} + i\epsilon_{ab}\right )
= e_a^*e_b
\label{fact}
\end{equation}
where $e_a$ is a complex covariant vector  with the property
\begin{equation}
\epsilon^{ab}e_a^*e_b = i.
\end{equation}
Note that this factorization has a $U(1)$ ambiguity under
$e_a \mapsto e_a \exp i\phi$.
A dimensionless complex coordinate can be defined by
\begin{equation}
z = e_ax^a/\surd 2 \ell_B.
\end{equation}
The coherent state condition is  then
\begin{equation}
b|\Psi_n(\bar {\bm x}, \tilde g)\rangle =   \bar z|\Psi_n(\bar {\bm x}, \tilde g)\rangle,
\end{equation}
with
\begin{equation}
\bar z  = e_a\bar x^a/\surd
2\ell_B.
\end{equation}

A common choice of electromagnetic gauge is the so-called ``symmetric
gauge'',  which derives from the factorization (\ref{fact}) of the Euclidean metric
$\delta_{ab}$ in terms of $(e_1,e_2)$ =  $2^{-\frac{1}{2}}(1, i)$.
Then the symmetric gauge is given by
\begin{equation}
A_a(\bm x) =  2^{-\frac{1}{2}} i (z^*\bm e_a - z\bm e_a^*).
\label{symg}
\end{equation}

For general $\tilde g_{ab}$ with a factorization (\ref{fact}),  (\ref{symg}) is still a valid gauge choice  
$A_a(\bm x, \tilde g)$  that is now ``symmetric''  with respect
to the unimodular metric $\tilde g_{ab}$ = $e^*_ae_b + e_a e^*_b$.
In this gauge, the Schr\"odinger representation of the guiding-center
harmonic oscillator
operators  $b$ and $b^{\dagger}$ is
\begin{equation}
b = \textstyle{\frac{1}{2}}z^* + \frac{\partial}{\partial z}, \quad
b^{\dagger} = \textstyle{\frac{1}{2}}z  - \frac{\partial}{\partial
                z^*}.
\end{equation}
where
$\partial f(z^*)/\partial z$ = $\partial f(z)/\partial z^*$ = 0.
The corresponding representation of the Landau-orbit harmonic oscillator operators is
\begin{equation}
\label{eq:complex_repr}
a = {\textstyle\frac{1}{2}} z + {\textstyle\frac{\partial}{\partial
  z^*}}, \quad a^{\dagger} = {\textstyle\frac{1}{2}} z^* -
  {\textstyle\frac{\partial}{\partial z}}.
\end{equation}

This immediately implies that the Schr\"odinger wavefunction 
$\Psi_n(\bm x; \bar{\bm x}, \tilde g)$ $\equiv$
$\langle \bm x|\Psi_n(\bar{\bm x}, \tilde g) \rangle$
of the
coherent state  has the functional form
\begin{equation}
\Psi_n(\bm x;\bar {\bm x}, \tilde g) 
= f_n(z^* -\bar z^*) e^{-\frac{1}{2} (\bar z^*\bar z - 2 \bar z^*  z +
  z^*z)},
\label{cohll}
\end{equation} 
where $f_n(z^*)$ = $(-1)^nf_n(-z^*)$ is a holomorphic function of
$z^*$.
Here $(-1)^n$ is the parity of the Landau-orbit coherent state with respect to
spatial inversion around the point $\bar x$ at which it is centered.

A continuous rotational invariance with respect to the metric $\tilde
g_{ab}$ exists if
\begin{equation}
[H, L(\tilde g)] = 0, \quad L(\tilde g) =
 - {\textstyle\frac{1}{2}}\hbar eB  \tilde g^{ab}\gamma_{ab}.
\label{genrot}
\end{equation}
In this case,  the function $f_n(z^*)$ is the homogeneous polynomial
\begin{equation}
f_ n(z^*) =  \frac{(z^*)^n}{\surd (n!)} ,
\end{equation}
and an {\textit{arbitrary} lowest-Landau level ($n$ = 0) wavefunction  has the
well-known form
\begin{equation}
\Psi_{0\alpha}(\bm x) =   F_{\alpha}(z) e^{-\frac{1}{2}  z^*z},
\label{lll}
\end{equation}
where $F_{\alpha}(\bm z)$ is a holomorphic function.   This is a
familiar result in the rotationally-symmetric case  $\tilde g_{ab}$ =
$\delta_{ab}$ (when $z^*z$ = $\frac{1}{2}\delta_{ab}x^ax^{*b}/\ell_B^2$), but remains true
if (\ref{genrot}) holds for \textit{any} $\tilde g_{ab}$, when there is a
``generalized rotational symmetry''.   

However, such a continuous
symmetry is \textit{not} a generic property of electrons moving inside
condensed matter, and 
the holomorphic lowest-Landau-level  property (\ref{lll}) disappears
in the absence of  this 
continuous symmetry.     Starting with the Laughlin
wavefunction, many models for the FQHE have incorporated this holomorphic lowest-Landau-level
structure, and it has often been regarded as a fundamental
ingredient of a theory of the FQHE.   However, (as evidenced by the
fact that a Laughlin-like FQHE state also  occurs in the second Landau
level) the holomorphic structure (\ref{lll}) can only have ``toy
model'' status as a convenient simplification, and thus \textit{can play no fundamental role
in a generic theory of the FQHE}.

In contrast, the holomorphic
coherent-state property
(\ref{cohll}) is quite generic, with no requirement for a  continuous rotational symmetry.
In the generic absence of such a symmetry, the functions $f_n(z)$ are
\textit{holomorphic, but not polynomial}. 
The coherent state has the structure
\begin{equation}
|\Psi_n( \bm x, \tilde g)\rangle 
=
f_n(a^{\dagger}) |\bar {\bm x},\tilde g\rangle, 
\end{equation}
where
\begin{equation}
a|\bar {\bm  x},\tilde g\rangle = 0, \quad
b|\bar {\bm  x},\tilde g\rangle = 
\bar z |\bar {\bm  x},\tilde g\rangle.
\end{equation}

 Holomorphic functions are
characterized by their zeroes (plus their asymptotic
behavior for large $z$).
However, the location of
the zeroes of the functions $f_n(z^*)$ will vary continuously  with the so-far arbitrary
choice of the coherent state metric $\tilde g_{ab}$ = $\bar g_{ab}$ (apart from the
generic occurrence of a zero at $z^*=0$ if $n$ is odd, and its absence if
$n$ is even, as a consequence of momentum-space inversion symmetry
(\ref {invert1})).

To use
$f_n(z^*)$ for a
standardized description, a ``natural choice'' of metric must be uncovered.
The key to this will be a property of the Landau levels  that is now generally called their ``Hall
viscosity'', first discussed in the context of 2D  Landau levels 
by Avron, Seiler and Zograf (ASZ)\cite{ASZ}.  This will however need a
significant reformulation of the Hall viscosity in the absence of a continuous rotational symmetry.

\section{ Hall viscosity and the Landau-orbit metric}
\label{sec:viscosity}
The viscosity of a continuous fluid is the usually-dissipative relation between
the stress (the current-density of momentum) in a fluid and the
gradient of its flow velocity.    
The use of a formalism with consistent index-placement that
distinguishes between covariant and contravariant indices will be
important in clarifying the structure of the stress tensor when
rotational
invariance is not present.

Momentum, like a force, is a covariant vector, and  the momentum
density  $T^0_{\;a} (\bm x,t)$ is locally conserved in translationally-invariant continuum
dynamics.   If translational invariance is broken by coupling to
background fields, the continuity relation is
\begin{equation}
\partial_t T^0_{\;a} + \partial_bT^b_{\; a} =  F_a,
\label{contin}
\end{equation}
where $T^a_{\; b}(x,t)$ is the mixed-index stress tensor, and
$F_a(x)$ is the body force exerted on the continuous medium by the
background fields that break translational invariance.

As a mixed-index tensor, there is in general no symmetry relating the
two indices.    However, if the system has a rotational invariance
under coordinate transformations 
that leave the Euclidean metric tensor $\delta_{ab}$ invariant, the
stress tensor has the Cauchy symmetry property
\begin{equation}
\delta_{ac}T^c_{\; b} -  \delta_{bc}T^c_{\; a} = 0.
\label{Cauchy}
\end{equation}

When there is such a rotational invariance, it is well-known that
it is possible to obtain the expression for the 
stress tensor from the variation of the action 
with respect to the metric that defines the invariance
\begin{equation}
\delta S = -{\textstyle\frac{1}{2}}\int d^4x \surd
              (-g)T_{\mu\nu}\delta g^{\mu\nu}.
\label{varymetric}
\end{equation}
The model for this is the derivation of the 
stress-energy tensor   $T^{\mu}_{\;\nu}$  in General Relativity (GR), where (local) rotational invariance
generalizes to  (local) Lorentz invariance. 
and the Cauchy relation becomes
\begin{equation}
g_{\mu\sigma}T^{\sigma}_{\;\nu} - T^{\sigma}_{\;\mu}g_{\sigma\nu} = 0,
\label{Lorentz}
\end{equation}
where $g_{\mu\nu}$ is the gravitational metric.

The relation (\ref{varymetric})  (or its dual where $g_{\mu\nu}$ is
varied) is often taken as the fundamental definition of the stress-energy tensor,
given in the form $T_{\mu\nu} = g_{\mu\sigma}T^{\sigma}_{\; \nu}$ or
$T^{\mu\nu}$ = $T^{\mu}_{\;\sigma}g^{\sigma\nu}$, but this is a symptom
of a common practice in GR  to hide dependence on the metric by using it to raise or
lower indices away from their ``natural'' positions (\textit{e.g}. by
forming metric-dependent constructions such as 
$F^{\mu\nu}$ = $g^{\mu\sigma}g^{\nu\tau}F_{\sigma\tau}$).

However, on a fundamental level, stress (or stress-energy) is the
response to \textit{strain} of the matter fields in space (or
space-time), and what is arguably its true defining relation is the variation of the action
under the infinitesimal \textit{active} diffeomorphism of the matter
fields  $\varphi(x)$ $\mapsto$ $\varphi (x')$ = $\varphi(x + \delta u(x))$, where
$\delta u^{\mu}(x)$ = $ u^{\mu}(x) \delta\lambda$.
This  definition of the stress-energy tensor is
\begin{equation}
\delta S = \int d^4x \surd (-g)T^{\mu}_{\; \nu} \nabla_{\mu}\delta u^{\nu}.
\label{diff}
\end{equation}
The change in the inverse metric under the diffeomorphism
(\textit{i.e.}, after the pull-back of the coordinate system to match
the diffeomorphism) is
\begin{equation}
g^{\mu\nu}(x) \mapsto g^{\mu\nu}(x') + \delta g^{\mu\nu}(x')
\end{equation}
with
\begin{eqnarray}
\delta g^{\mu\nu} &=& 
\mathcal L_u (g^{\mu\nu} ) \delta \lambda 
\nonumber \\
&=& -(g^{\mu\sigma}\nabla_{\sigma} \delta u^{\nu} +g^{\nu\sigma}\nabla_{\sigma}
 \delta    u^{\mu}) ,
\end{eqnarray}
where $\mathcal L_u(\ldots )$ is the Lie derivative.
  Then the combination of the fundamental definition
(\ref{diff}) with the local Lorentz-invariance property (\ref{Lorentz})
leads to the familiar result (\ref{varymetric}).

The QHE is exhibited by  incompressible
2D electron fluids.   In the case of the integer QHE, this results
from the Pauli principle, where (in the zero-temperature limit) all
Landau levels are either filled or empty.    In this case, there are
no gapless collective degrees of freedom of the electron gas that 
can transmit forces through the  bulk, or allow dissipation of energy.

This means that the gapped quantum incompressible liquid is quite
different from the  notional  classical incompressible liquid, 
which is just the limiting case of a classical  liquid with a large bulk modulus.
The non-relativistic classical fluid supports sound waves (pressure waves), which
travel with a group velocity that diverges in the limit in which the
bulk modulus becomes infinite.    (This of course violates causality
as the speed of sound eventually exceeds the Lorentz speed $c$, so the ``thought
experiment'' of taking the incompressible limit of a classical
non-relativistic fluid cannot in principle be realized.)  Formally,
the classical incompressible fluid supports a hydrostatic pressure which is
instantaneously spatially-uniform throughout the fluid,  so the trace of the
stress tensor is spatially homogeneous:
\begin{equation}
T^a_{\; a} (\bm x,t) = p(\bm t)\delta^a_a.
\end{equation}
However the gapped quantum incompressible fluid does not support sound
waves, and thus has a \textit{traceless} stress tensor:
\begin{equation}
T^a_{\; a}(\bm x,t) = 0.
\end{equation}

The viscosity tensor is the linear relation between the stress in a
fluid and a non-uniform flow velocity field $\bm v(\bm x,t)$
= $v^a\bm e_a$
that breaks translational invariance.    It is also fundamentally a
locally-defined 
mixed-index tensor
\begin{equation}
\label{eq:viscosity}
T^a_{\;b} (\bm x,t) = \eta^{a\;c}_{\;b\;d}(\bm x,t) \partial_cv^d(\bm x,t) +
O( v^2) .
\end{equation}
In the zero-temperature limit, the viscosity of the gapped
incompressible fluid must be dissipationless, which is the
antisymmetry
condition 
\begin{equation}
\eta^{a\;c}_{\;b\;d} = - \eta^{c\;a}_{\;d\;b}.
\end{equation}
The dissipationless viscosity must also be odd under time reversal.

In two dimensions, the tracelessness and antisymmetry of the viscosity
tensor
that is implied by quantum incompressibility 
means that it must have the structure 
\begin{eqnarray}
\eta^{e\;f}_{\;b\;d} &=& \epsilon^{ae}\epsilon^{cf} \eta^H_{abcd}, \\
\eta^H_{abcd} &=& \eta^H_{bacd} = \eta^H_{abdc} = -\eta^H_{cdab} \\
&=& {\textstyle \frac{1}{2}} \epsilon\left (\epsilon_{ac}\eta^H_{bd} 
+ \epsilon_{ad}\eta^H_{bc} 
+ \epsilon_{bc}\eta^H_{ad} 
+ \epsilon_{bd}\eta^H_{ac} \right ). 
\end{eqnarray}
where $\epsilon$ $\equiv$  $\epsilon_{12}$
is the Pfaffian of the 2D Levi-Civita symbol.
Here the rank-4 tensor  $\eta^H_{abcd}$ is  odd under time-reversal, and 
 $\eta^H_{ab}(\bm x,t)$ is a symmetric rank-2 covariant tensor that 
is also odd under time-reversal.    This  rank-2 tensor will here be called the  2D 
``Hall viscosity'' tensor, and satisfies the 
stability condition 
\begin{equation}
\det \eta^H \ge 0. 
\end{equation}

If there is rotational invariance, $\eta^H_{ab}$ must be proportional
to the Euclidean metric $\delta_{ab}$.    Previous work by other
authors on the Hall
viscosity has generally specialized to this case, with the correspondence
\begin{equation}
\eta^H_{ab} =  \bar\eta^H \delta_{ab}.
\end{equation}
In particular,  Read\cite{Read} has provided an interpretation of  the
time-reversal-odd scalar $\bar\eta^H$ as
\begin{equation}
\bar \eta^H= {\textstyle\frac{1}{2}} \hbar n_e \bar s =
\textstyle\frac{1}{2} \bar l
\end{equation}
where $n_e$ is the 2D electron density, and $\bar s$ is an ``intrinsic
orbital spin per electron'', or, equivalently, that $\bar l$ is the (local)
density of ``intrinsic orbital angular momentum'' of the electron fluid, a quantity that only has
meaning if there is a continuous rotational symmetry.  Equivalently,
Read's  formula  can be stated as the result that, when there is such a
rotational symmetry, 
\begin{equation}
\delta^{ab}\eta^H_{ab}     = \bar l.
\end{equation}

To explicitly calculate the stress tensor, we must first obtain its generic expression in the absence of rotational symmetry using the fundamental definition (\ref{diff}). Similar derivation, though without consistent index placement, can be found in the reference.\cite{Bradlyn} We write down the action for 2D electron fluids in $U(1)$ gauge fields $A_0$ and $A_a$,
\begin{align}
S &= \int d^2x dt\, \mathcal L, \\
\mathcal L &= \frac{i}{2}\hbar\psi^\dagger\partial_0 \psi - \frac{i}{2}\hbar(\partial_0 \psi^\dagger)\psi + eA_0 \psi^\dagger\psi - \mathcal H
\end{align}
where we have assumed that the underlying manifold is the 2D Euclidean plane, i.e., $g_{ab}=\delta_{ab}$ and $\sqrt g=1$. For a system with Galilean invariance the Hamiltonian density $\mathcal H$ is just
\begin{align}
\mathcal H &= \frac{\hbar^2}{2}m^{ab} (D_a\psi)^\dagger D_b \psi, \\
D_a&= \partial_a - i\frac{e}{\hbar}A_a,
\end{align}
where $m^{ab}$ is the inverse mass tensor induced by the band structure, and therefore has nothing to do with the (inverse) manifold metric $g^{ab}$. For the purpose of generality we leave the form of the Hamiltonian density unspecified. Under the infinitesimal spatial diffeomorphism $x'^a=x^a+\delta u^a$ the $U(1)$-covariant derivatives transform as
\begin{align}
D'_{a} = (\delta^b_a-\partial_a \delta u ^b) D_b,
\end{align}
and the variation of the action is
\begin{align}
\delta S &= \int d^2xdt \, T^a_{\;b} \partial_a \delta u^b, \\
T^a_{\;b} &= \frac{\delta \mathcal H}{\delta (D_a \psi)^\dagger} (D_b \psi)^\dagger +  \frac{\delta \mathcal H}{\delta (D_a \psi)} D_b \psi + \mathcal L \delta^a_b.
\end{align}
For the IQHE, which is essentially a single-particle problem, the stress tensor can be written in terms of single-particle operators,
\begin{align}
T^a_{\;b} &= \frac{1}{2\pi l_B^2} \tau^a_{\;b},\\
\tau^a_{\;b} &= \frac{1}{2} \{ v^a, p_b\},  \quad v^a =\frac{\partial \varepsilon(\bm p)}{\partial p_a},
\end{align}
where $\varepsilon(\bm p)$ is the single-particle Hamiltonian. The traceless part of the stress tensor can also be obtained
\begin{align}
	\tilde \tau^a_{\;b} = \tau^a_{\;b} - \tfrac{1}{2} \tau^c_{\;c} \delta^a_{\;b} = -i\epsilon^{ac} [\gamma_{bc}, \varepsilon(\bm p)],
\end{align}
where the $\gamma_{ab}$ operators are defined in Section (\ref{sec:algebra}).

Now we are ready to calculate the response of the stress tensor to a spatially-uniform but time-varying shear strain $\partial_a \delta u^b(t) = \lambda_a^b(t)$. The perturbative Hamiltonian is
\begin{align}
\delta \varepsilon (t) = -\tau^a_{\;b} \lambda_a^b(t) = -\tilde \tau^a_{\;b}  \lambda_a^b(t).
\end{align}
According to linear response theory, the induced stress is determined by the retarded stress-stress correlation function,
\begin{align}
\tilde \tau^a_{\; b} (t) &= \int dt' G^{a\;c}_{\;b\;d}(t-t')  \lambda_c^d(t'),\\
G^{a\;c}_{\;b\;d} (t) &=  \sum_{n}\frac{\nu_n}{\hbar}i\Theta (t)\langle n | [\tilde \tau^a_{\;b} (t), \tilde \tau^c_{\;d}(0)] |n \rangle,
\end{align}
with $\Theta(t)$ being the step function and $\nu_n$ = 1 if Landau level $n$ is filled and
$\nu_n$ = 0 if it is empty. We then Fourier-transform it into frequency space, 
\begin{align}
G^{a\;c}_{\;b\;d} (\omega) &=  \int dt \, e^{i(\omega+i\eta) t} G^{a\;c}_{\;b\;d}(t),
\end{align}
where $\eta$ is an infinitesimal positive number. To obtain the formula for the Hall viscosity, what we need is the coefficient of the first order term in the $\omega$-expansion,
\begin{align}
\frac{dG^{a\;c}_{\;b\;d}(\omega)}{d\omega}\bigg|_{\omega=0}&=\sum_{n,n'} \frac{\hbar\nu_n}{(E_n-E_{n'})^2} \left( \langle n | \tilde \tau^a_{\;b} | n' \rangle \langle n' | \tilde \tau^c_{\;d} |n\rangle \right. \notag \\
&\left.- \langle n | \tilde \tau^c_{\;d} | n' \rangle \langle n' | \tilde \tau^a_{\;b} |n\rangle \right) \\ 
&=\sum_{n} \hbar\nu_n \epsilon^{ae} \epsilon^{cf}  \langle n | [\gamma_{be}, \gamma_{df}] | n \rangle.
\end{align}
Comparing it to the definition of the viscosity (\ref{eq:viscosity}), we arrive at the expression for the rank-4 tensor $\eta^H_{abcd}$,
\begin{align}
\eta^H_{abcd} = - \frac{i \hbar}{2\pi l_B^2} \sum_{n} \nu_n \langle n |[\gamma_{bc}, \gamma_{de}] |n\rangle.
\end{align}
Recall the commutator of the $\gamma_{ab}$ operators (\ref{Lie}), with which we obtain the expression for the 2D Hall viscosity tensor $\eta^H_{ab}$,
\begin{align}
\eta^H_{ab} = \frac{\hbar}{2\pi l_B^2}\sum_{n}\nu_n \langle n | \gamma_{ab} |n\rangle.
\end{align}

Based on this calculation, in each Landau level we can define a metric $g^n_{ab}$ by
\begin{equation}
\langle n|\{p_a,p_b\}|n\rangle = 
-4\pi \hbar^2 s_n {\rm sgn (eB)} g^n_{ab}
\label{llmetric}
\end{equation}
where 
\begin{equation}
s_n = -{\rm sgn }(eB) (n + {\textstyle\frac{1}{2}})
\end{equation}
is a \textit{topological spin} of the Landau level 
that appears in the semiclassical formula (\ref{semiclassical})
and  can also be obtained by inspection of the Landau-level
coherent state calculated with $\tilde g_{ab}$ $\propto$ $g^n_{ab}$
where $g^n_{ab}$ is the (non-unimodular)  Euclidean-signature metric
of a dimensionless distance measure associated with the Landau level.
Note that while the ``topological spin''  $s_n$ coincides with what Read has
called the ``intrinsic orbital angular momentum'' of a
rotationally-invariant Landau level, it is a topological invariant
with an existence independent of the presence or absence of a
rotational symmetry, described in more detail later.

The Landau-orbit contribution to the Hall viscosity is then
\begin{equation}
\eta^H_{ab} = \frac{\hbar}{2}\sum_n \nu_n s_n g^n_{ab}.
\end{equation}
For a rotationally-invariant
system, all Landau levels have the same metric
\begin{equation}
g_{ab} = \frac{|eB|}{2\pi \hbar}\delta_{ab}.
\end{equation}

We have now identified a ``natural'' unimodular  metric  $\tilde
g^n_{ab}$ $\propto$ $g^n_{ab}$ associated with
each Landau level, that allows the definition of a standard
localized Landau-orbit coherent state.

\section{Zeroes of the Landau-orbit wavefunction and topological spin}
\label{sec:zeroes}

The identification of an intrinsic unimodular metric $\tilde g^n_{ab}$
associated with the shape of  the Landau orbit in a given Landau level
now allows a systematic analysis of its wavefunction.

A basis of states in momentum space is defined by 
\begin{equation}
{\textstyle\frac{1}{2}}\tilde g^{ab}p_ap_b|n, \tilde g\rangle = \hbar |eB| (n
+{\textstyle\frac{1}{2}})|n, \tilde g\rangle.
\label{trun}
\end{equation}
where $\tilde g_{ab}$ is some unimodular metric.
Assuming that the expansion (\ref{ham}) is truncated at some finite
maximum $n$ = $n_{\rm max}$,  there are only non-vanishing  matrix elements 
$\langle n_1|H|n_2\rangle$
for $|n_1-n_2| \le 2n_{\rm max}$.
An arbitrary choice of $\tilde g_{ab}$ can initially be made, and the
problem can be diagonalized in a basis truncated  to $n \le N$ for some
suitably-large $N$.  For each eigenstate of interest, the metric
$g^n_{ab}$ (\ref{llmetric}) is calculated, and the problem is
rediagonalized in a basis (\ref{trun}) with $\tilde g_{ab}$ = $\tilde g^n_{ab}$$\propto$
$g^n_{ab}$, and a suitably large truncation (this process can be
iterated if necessary).    A Taylor-series expansion that approximates
$|\psi_n\rangle$ by
\begin{equation}
|\psi_n\rangle =  f_n(a^{\dagger}) |0,\tilde g^n\rangle
\end{equation}
where $f_n(z^*)$ is a finite-degree polynomial is then obtained, and
the zeroes of the polynomial  can be obtained with a standard root
finder.     We studied models with quartic dispersions, $2n_{\rm max}$
= 4, and used MPACK\cite{MPACK}, a version
of LAPACK adapted to use the GMP multiprecision library to get very
accurate eigenstates, and studied how the patterns of  zeroes varied with the
truncation $N$.  We observed that  the polynomial roots $\{z_i\}$ converged to
stable values inside some radius in the complex plane (which is also a
representation of the momentum plane)   which grew as $N$ was
increased.    The use of MPACK allowed accurate determination of the
zeroes for large $N$ of  order 1000, exposing them in a large ``window"
in momentum space.

The polynomial form
\begin{equation}
|\Psi |^2 \propto \exp V, \quad
V(z,z^*) = - |z|^2 + \sum_i \ln |z-z_i|^2 
\end{equation} 
gives the amplitude of the Landau-orbit wavefunction in momentum space
in terms of a potential that is a solution of Poisson's equation with a
Laplacian defined by the metric $\tilde g^n_{ab}$, with a uniform
positive charge inside a circular disk of large radius (that fixes the
asymptotic form of the holomorphic function $f_n(z^*)$), and unit
negative point charges at the zeroes.

For a strictly-monotonic dispersion $\varepsilon (\bm p)$,
The semiclassical quantization identifies the $n$'th quantized Landau
orbit with a region of momentum space with total area $2\pi \hbar |eB|$
between the closed contour
$\mathcal A(E^-)$ = $2\pi n\hbar |eB|$ and $A(E^+)$ = $2\pi (n+1)\hbar
|eB|$.   This region has the topology of a disk  (Euler characteristic
$\chi $ = 1) for $n$ = 0, or an annulus (Euler characteristic
$\chi$ = 0) for $n > 0$, and can be thought of as a ``fattened''
semiclassical orbit.
 When $f_n(z^+)$ is obtained using the metric $g^n_{ab}$, we
observed that it is possible to choose a threshold $c$ so the region
defined by $|\Psi_n(z,z^*)|^2 $ =  $|\Psi_n(\bm p)|^2 > c$  covers an
area of order $2\pi \hbar |eB|$, and defines a
zero-free region with the same topology as the semiclassical
prediction that is topologically a disk for $n$ = 0, 
and for $n > 0$ an annulus that encircles $n$ zeroes, hence defining a ``topological spin" $s_n=n+\frac{1}{2}$ of the Landau level. It is also possible to choose $c$ so most of the weight of the wavefunction is within the region.

This picture remains true when $f_n(z^*)$ is calculated using a 
different metric that is  close to $\tilde g^n_{ab}$, but eventually breaks down for a
choice of metric sufficiently different from $\tilde g^n_{ab}$: the
pattern of zeroes changes with the metric, and  for $n > 0$
the central zeroes  eventually begin to  ``leak out'' of the annulus defined by
the ``fattened'' semiclassical orbit.

When a guiding-center coherent state of the Landau orbit is
constructed using  $\bar g_{ab}$ = $\tilde
g^n_{ab}$,  there is also a mapping from the complex
plane
to real space, with the origin at the guiding center of the orbit, and the
zeroes of $f(z^*)$ become the zeros of the coherent state wavefunction.

The analysis of the generic Landau orbits has now  identified a
topological spin $s_n$ and a non-unimodular  metric
$g^n_{ab}$ as well as its unimodular version $\tilde g^n_{ab}$, which in the rotationally-invariant case reduces to $\tilde g^n_{ab}$ = $\delta_{ab}$.

\section{A case study: the quartic dispersion}
\label{sec:casestudy}
\begin{figure}[]
\includegraphics[trim=100mm 5mm 100mm 15mm,clip,width=\linewidth]{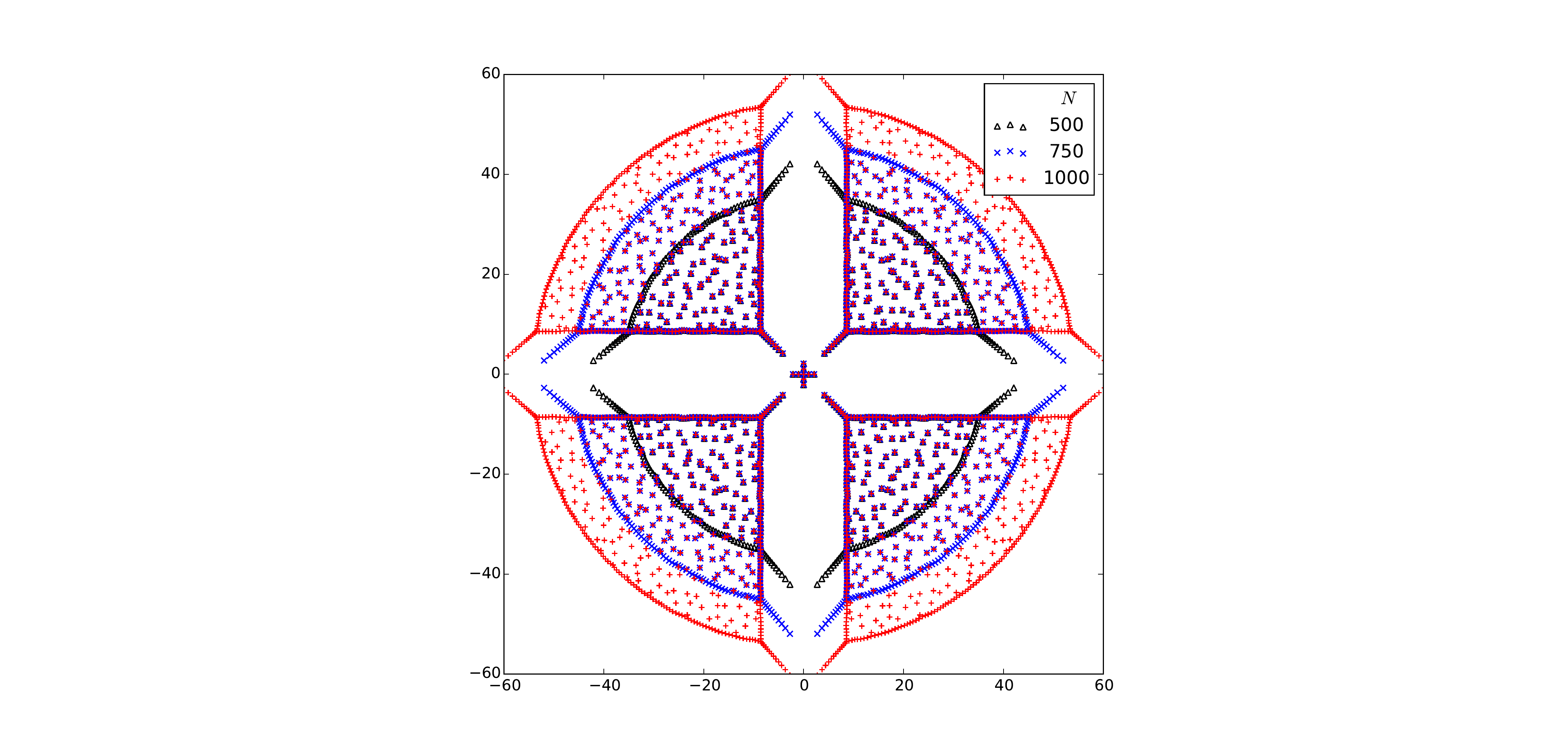}
\caption{Zero distribution of $f_{10}$ for class I quartic term with $c-1=4$ at different truncations $N$. Here $N$ is not the degree of the polynomial, but the number of fixed-parity states we kept for diagonalization. For each $N$, we used a sufficiently high precision so that further increasing the precision would not do any better in terms of the locations of the zeroes. As $N$ increases, the region covered by the zeroes expands. However, the existing pattern of zero distribution doesn't change, which justifies the validity of our results.}
\label{fig:cutoff}
\end{figure}

We will now present a detailed case study of the simplest model
dispersion that exhibits the generic property that there is no
congruence between the shapes of different semiclassical orbits.

This is the model with quadratic and quartic terms in the dispersion:
\begin{equation}
  H =  A_1^{ab} p_ap_b   +
{\textstyle\frac{1}{4}}A_2^{abcd}\{p_a,p_b\}\{p_c,p_d\}.
\end{equation}
Up to the addition of the quadratic Casimir, which is a constant, and
does not affect the semiclassical orbits, there are 
four possible classes of quartic terms 
compatible with strict monotonicity.   If $\tilde g_i^{ab}$ are the
inverses of unimodular metrics:
\begin{eqnarray}
\nonumber
\tilde A_{2,\rm I }^{ab,cd} &=&  \tilde g_1^{ab}\tilde g_2^{cd} + \tilde g_2^{ab}\tilde
              g_1^{cd},\quad  {\textstyle\frac{1}{2}}(\tilde g_1^{ab}
                                + g_2^{ab}) = c \tilde g_0^{ab},
                                 \\
\nonumber
\tilde A_{2,\rm II}^{ab,cd} &=&  \tilde g_1^{ab}u_1^cu_1^d + u_1^au_1^b\tilde g_1^{cd} , \\
\nonumber
\tilde A_{2,\rm III}^{ab,cd} &=&   u_1^au_1^bu_2^cu_2^d + u_2^au_2^bu_1^cu_1^d,
                    \, \; u^a_1u^b_1 + u^a_2u^b_2 = \lambda \tilde g_0^{ab},\\
A_{2,\rm IV}^{ab,cd} &=&  u_1^au_1^bu_1^cu_1^d.
\end{eqnarray}
where $\lambda  > 0$, and $c \ge 1$.
Strict monotonicity also requires that the quadratic term $A_1^{ab}$ has
the form $\lambda' \tilde g'^{ab}$, $\lambda'\ge  0$, but the weaker condition that the spectrum
of H is stable (\textit{i.e.}, bounded below) only requires that
$\epsilon_{ac}\epsilon_{bd}A_1^{ab}u_i^cu_i^d \ge  0$ in classes II, III,
IV.

\begin{figure}[]
\includegraphics[trim=15mm 5mm 15mm 10mm,clip,width=\linewidth]{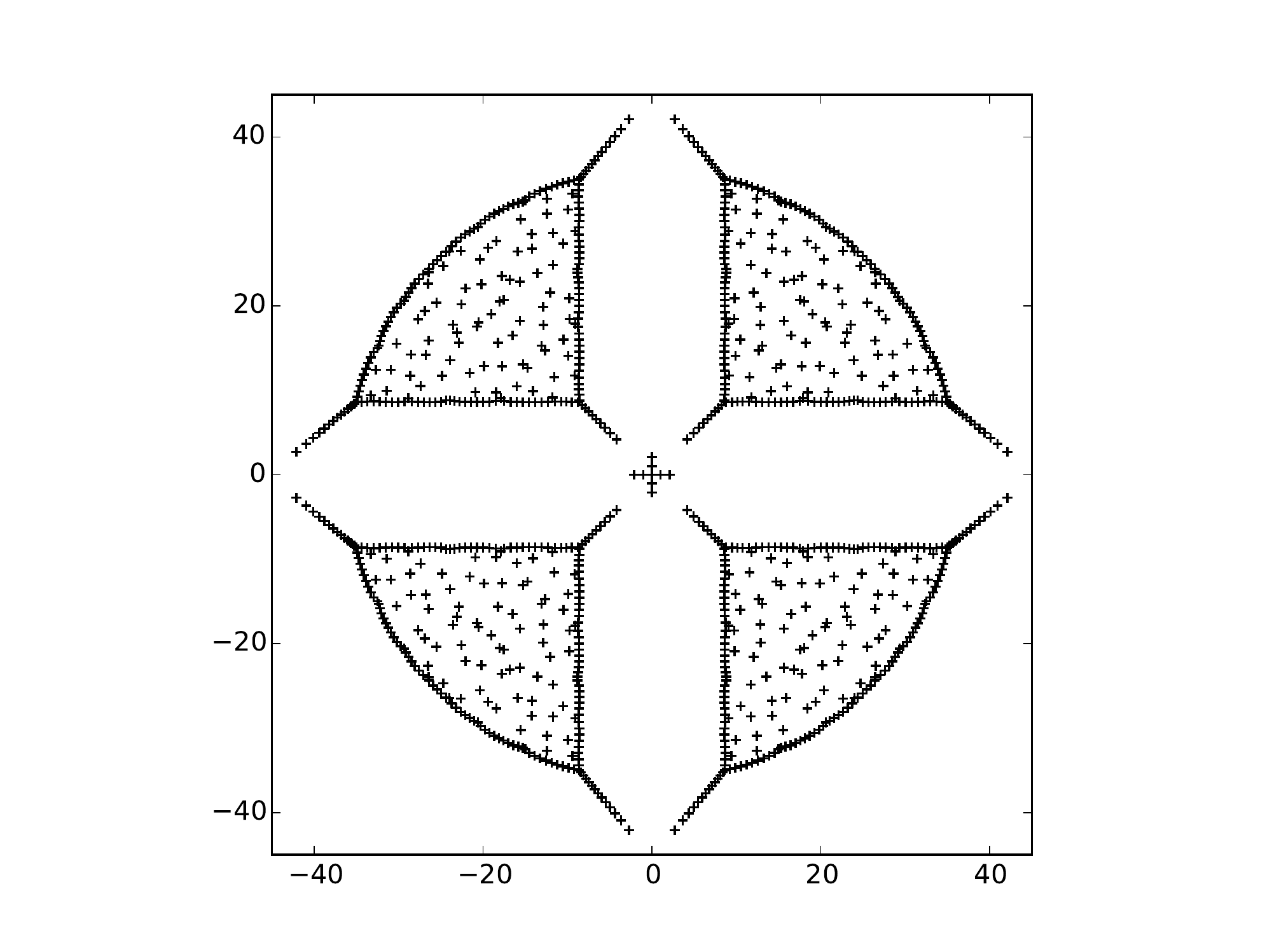}
\caption{Zeroes of $f_{10}$ for class I quartic term with $c-1=4$ at a truncation of 500 fixed-parity states. Ten central zeroes, with two being degenerate at the origin, form a cross due to the $C_4$ symmetry. The peripheral zeroes also have interesting features, such as four ``spikes" pointing to the origin and line charges with constant density that bound 2D regions of quasi-uniform distributions. The arcs on the boundary are generated due to finite truncation and should not be regarded as part of the zero pattern.}
\label{fig:class1zeroes}
\end{figure}
The class I and III quartic terms have a $C_4$ symmetry with inverse
metric
$\tilde g_0^{ab}$.   If $A_1^{ab}$ $\propto$ $g_0^{ab}$ the full
model also has this symmetry, and all Landau orbits have the  same metric
given by the inverse of $\tilde g_0^{ab}$.
If, in addition, $\tilde g^{ab}_1$ = $\tilde g_2^{ab}$ = $\tilde
g_0^{ab}$,  so $c$ = 1,
class I models have a continuous  $SO(2)$ symmetry.
 
We first review the  purely quadratic case, 
where the quartic term is absent, $A_2^{abcd}$ = 0.
In this case 
\begin{equation}
H = {\textstyle\frac{1}{2}}E_0 \tilde g^{ab}\gamma_{ab} , \quad E_0
> 0,
\end{equation} 
where $\tilde g^{ab}$ is the inverse of a unimodular metric $\tilde g_{ab}$ 
and there is a pseudo-Galilean invariance with $\delta_{ab}$ replaced
by $\tilde g_{ab}$, and $m_e$ replaced by $(E_0\ell_B^2/\hbar)^{-1}$.
However, the resemblance to the Newtonian model is superficial, as
$(E_0\ell_B^2)^{-1}$ and $\tilde g_{ab}$ can be spatially-varying
(and their product can have Gaussian curvature) on a
completely-geometrically-flat plane with a Cartesian coordinate
system.   In this non-generic case, the Landau-orbit metric is $\tilde
g_{ab}$ independent of the Landau index,  with all
the zeroes of the Landau-orbit coherent state at its  origin.

The next-simplest case is the purely-quartic class I model, which also has
the non-generic feature of a common metric $\tilde g_{ab}$ for all
Landau orbits, because of its hidden $C_4$ four-fold rotational symmetry.
The generic model with the strict-monotonicity property  is given by
\begin{equation}
H = {\textstyle\frac{1}{2}}E_0 \{ \tilde g_1^{ab}\gamma_{ab},\tilde
g_2^{cd} \gamma_{cd}\},
\end{equation}
where $\tilde g_1^{ab}$ and $\tilde g_2^{ab}$  are the inverses of two
distinct
unimodular metrics $\tilde g^1_{ab}$ and $\tilde g^2_{ab}$.
In this case 
\begin{equation}
{\textstyle\frac{1}{2}}\left (\tilde g^1_{ab} + \tilde g^2_{ab}\right
) = c\tilde g_{ab}, \quad c \ge 1,
\end{equation}
where $\tilde g_{ab}$ is the common unimodular  metric of all the Landau orbits.
After a $SL(2,R)$ transformation the Hamiltonian is put into the form
\begin{equation}
H = E_0 \left ( (\gamma_{11} + \gamma_{22})^2 +
  (c-1)\{\gamma_{11},\gamma_{22}\}\right )
\end{equation}
which has an explicit $C_4$ symmetry under $(p_1,p_2)$ $\mapsto$
$(-p_2,p_1)$, or
$\gamma_{11} \leftrightarrow
\gamma_{22}$.
In the harmonic oscillator representation with $[a,a^{\dagger}]$ = 1,
\begin{equation}
\gamma_{11}  =  {\textstyle\frac{1}{2}} (a+a^{\dagger})^2, \quad
\gamma_{22}  =  -{\textstyle\frac{1}{2}} (a-a^{\dagger})^2,
\end{equation}
and
\begin{equation}
|\psi_n\rangle = f_n(a^{\dagger})|0\rangle, \quad a|0\rangle = 0.
\end{equation}
The $C_4$ symmetry is then the  symmetry of $H$  under  $a^{\dagger}
\mapsto ia^{\dagger}$.

Using the complex representation (\ref{eq:complex_repr}) the eigen-problem
\begin{equation}
H|\psi\rangle = E|\psi\rangle
\end{equation}
can be written as a fourth order complex differential equation on the
antiholomorphic part of the wavefunction:
\begin{align}
&(\partial^4_{z^*} + \alpha_2(z^*) \partial^2_{ z^*} +
  \alpha_1(z^*) \partial_{ z^*} +\alpha_0(z^*)) f(z^*) = 0, \nonumber \\
&\alpha_2 = -(2+\frac{8}{c-1})z^{*2},\nonumber  \\
&\alpha_1 = -(4+\frac{16}{c-1})z^*,\nonumber  \\
&\alpha_0 = z^{*4}-1-\frac{2}{c-1}(1-\frac{E}{E_0}),
\end{align}
whose solutions are entire functions\cite{book} (antiholomorphic functions
over the whole complex plane), and  can be analyzed using Nevalinna
theory (see, \textit{e.g}, the book  Ref.\onlinecite{book}).
The solutions have a property that
their \textit{ordes} $\rho(f)$, defined as
\begin{align}
&\rho(f) = \limsup_{r \to \infty} \frac{\ln(\ln||f||_{\infty,B_r})}{\ln r}, \\
&||f||_{\infty,B_r}= \sup \{|f(z^*)|: z^* \in B_r\},
\end{align}
where $B_r$ is the disk with radius $r$, are less or equal than 2.
 The
functions  $1/f(z^*)$ (the inverses of the solutions)  are meromorphic functions and a
similar notion of order incorporating the singularities can also be
defined\cite{book}. In that case the fact that the orders are less or
equal than 2 basically means the number of roots of $f(z)$ (namely the
singularities of $1/f(z^*)$) inside $B_r$ can go at most as $r^2(\ln
r)^k$ for some $k$ when $r$ goes to infinity.      
We  remark that this property is not unique to the quartic case,
but valid for any finite-order expansion of the Hamiltonian, and
presumably is related to the property that $f(z^*)\exp -
\frac{1}{2}z^*z$ is a normalizable wavefunction.

\begin{figure}[]

\textbf{Class I: $c-1>2$}
\includegraphics[trim=100mm 5mm 100mm 15mm,clip,width=\linewidth]{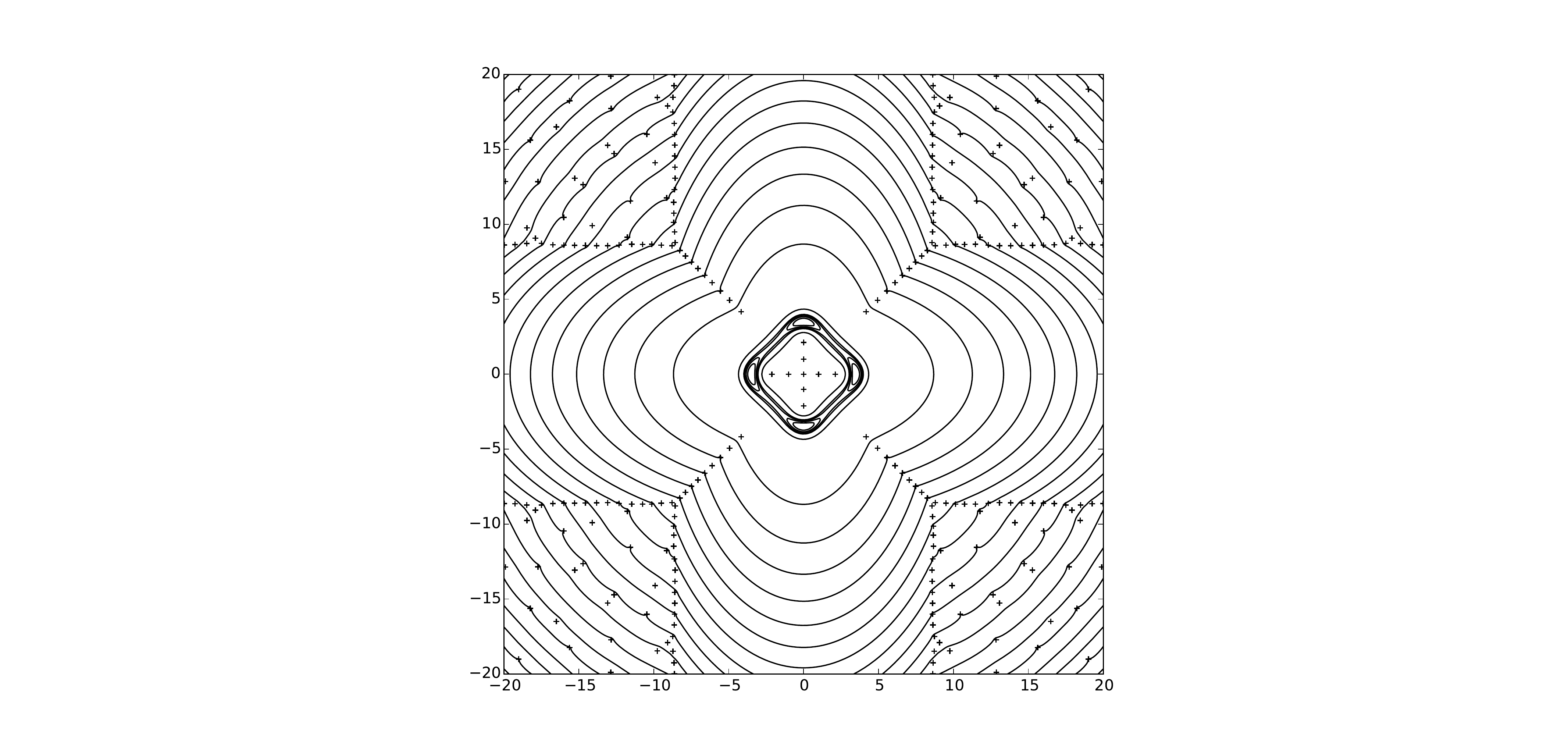}
\textbf{Class I: $c-1<2$}
\includegraphics[trim=100mm 5mm 100mm 15mm,clip,width=\linewidth]{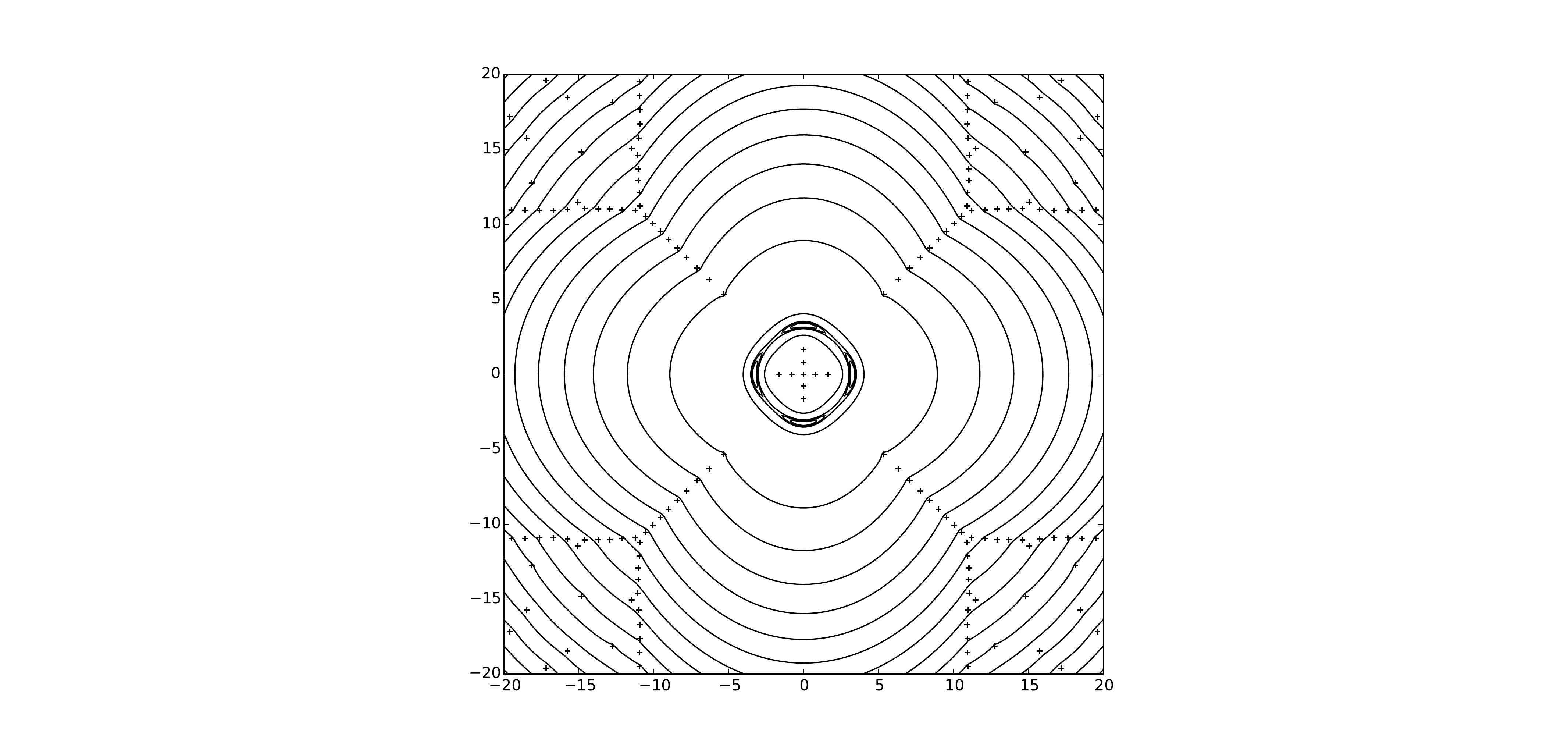}
\caption{Contour plots of $\ln |\Psi_{10}|^2$ for class I quartic term with $c-1 = 4>2$ and $c-1=1<2$ respectively. Both plots show piece-wise contours with the four spikes and the line charges as branch cuts. Another common feature is the existence of four maxima along the directions of the central cross and four saddle points along those of the spikes. Despite the similarities, the two plots also show qualitatively different shapes of the semiclassical orbits. We see that $c-1>2$ corresponds to concave shapes while $c-1<2$ to convex ones.}
\label{fig:class1contours}
\end{figure}

\begin{figure}[]
\textbf{Class II}
\includegraphics[trim=100mm 5mm 100mm 15mm,clip,width=0.9\linewidth]{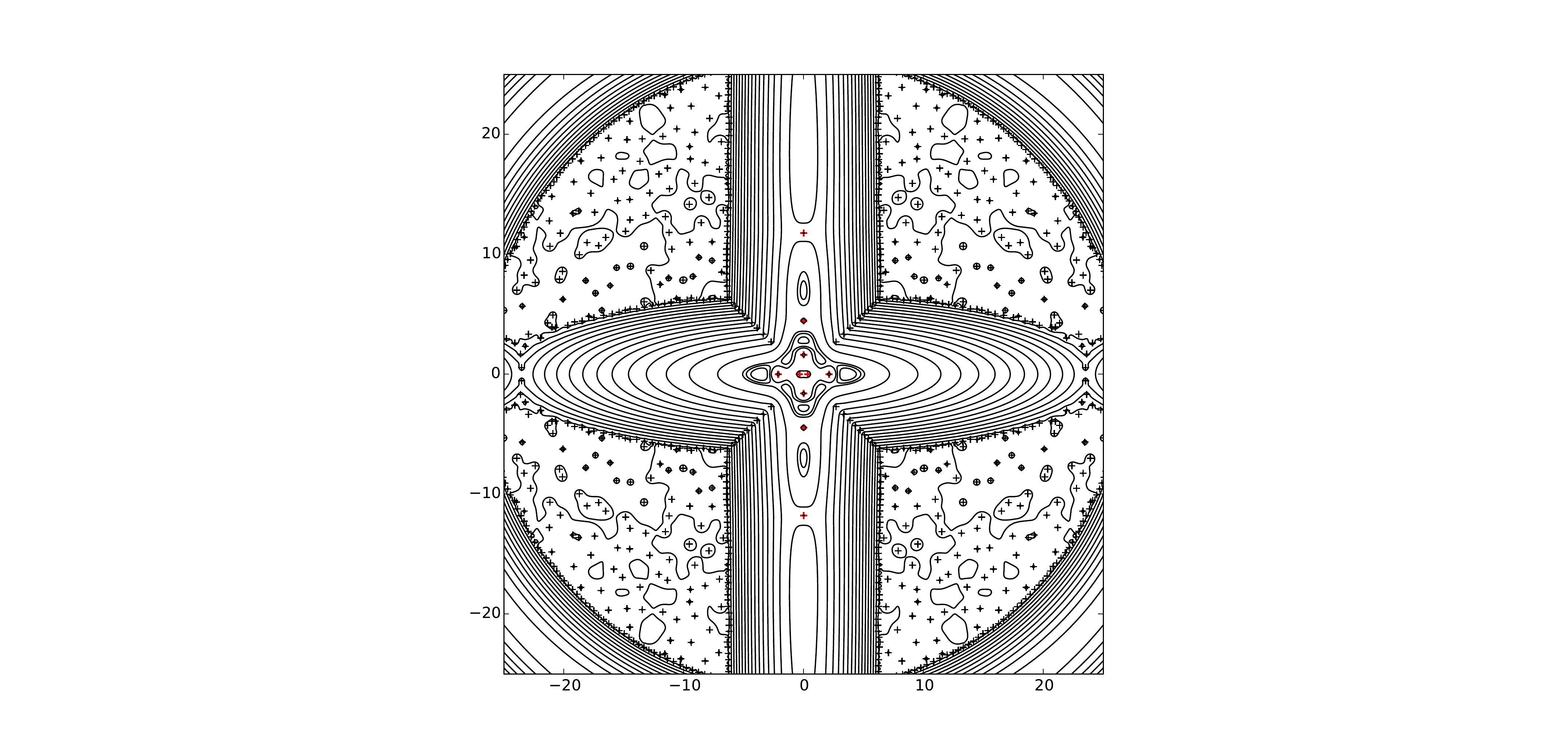}
\textbf{Class III}
\includegraphics[trim=100mm 5mm 100mm 15mm,clip,width=0.9\linewidth]{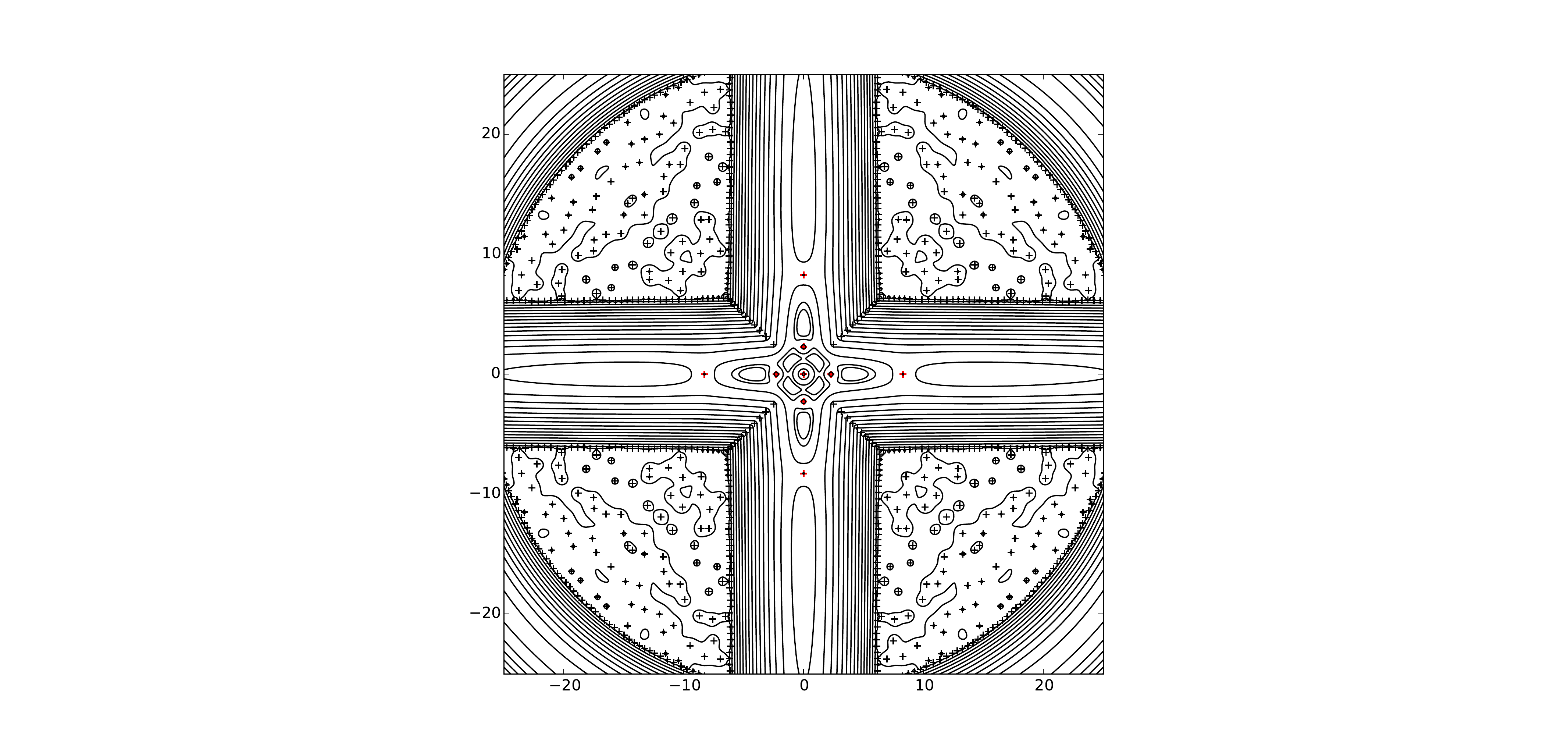}
\textbf{Class IV}
\includegraphics[trim=100mm 5mm 100mm 15mm,clip,width=0.9\linewidth]{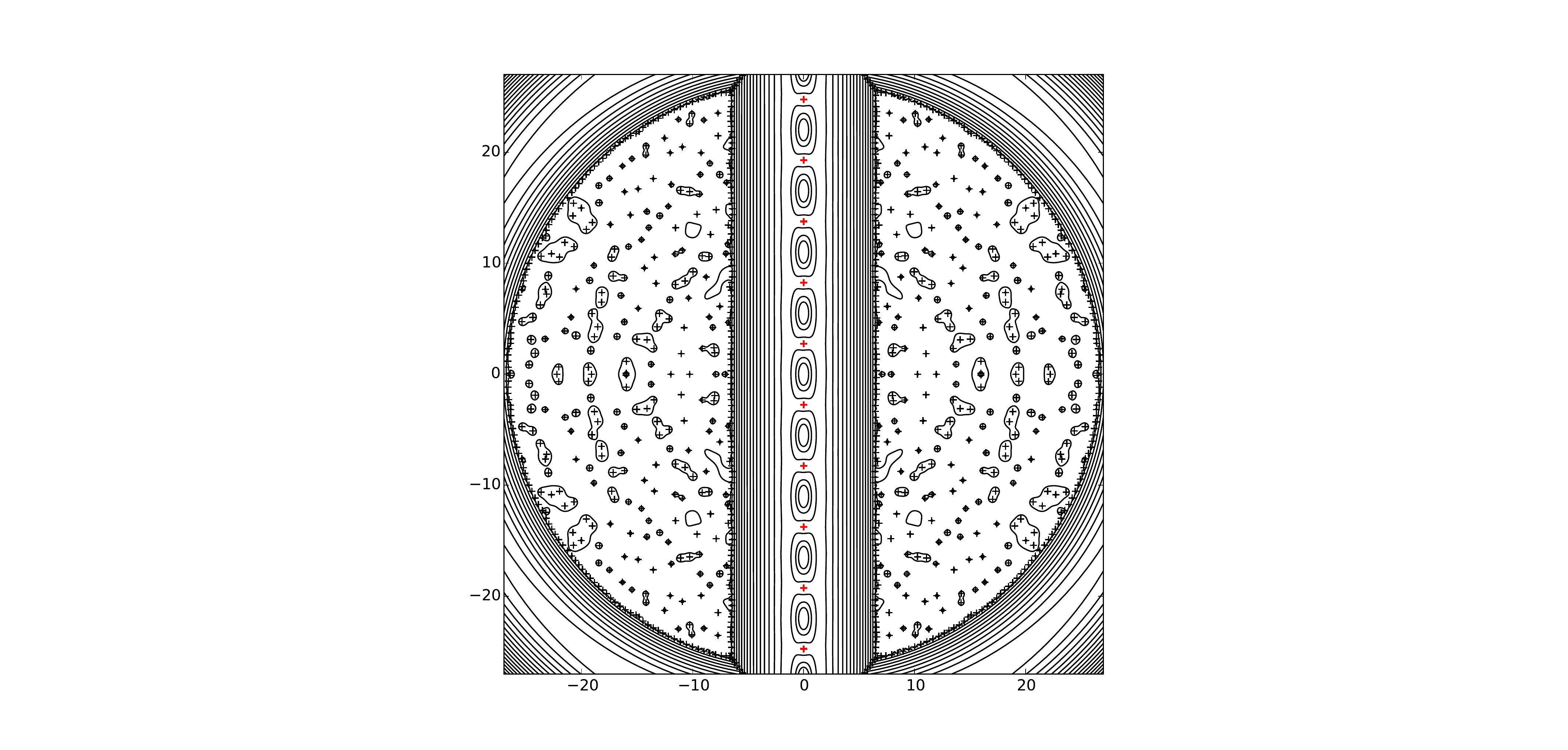}
\caption{Contour plots of $\ln |\Psi_{10}|^2$ for class II-IV quartic terms, with $\{p_x^2,p_x^2+p_y^2\}$, $\{p_x^2,p_y^2\}$, $p_x^4$ as their Hamiltonians respectively. Although classes II, III have closed contours that enclose part of the central zeroes (red pluses), none of the three has a clean separation between all the central zeroes and the rest ones. The contours inside the 2D-distribution regions also demonstrate qualitatively different features from class I: the puddle-like shapes indicate uniformly vanishing amplitudes.}
\label{fig:classes234}
\end{figure}

\begin{figure}[]
\textbf{$n=0$}
\includegraphics[trim=0mm 5mm 0mm 10mm,clip,width=\linewidth]{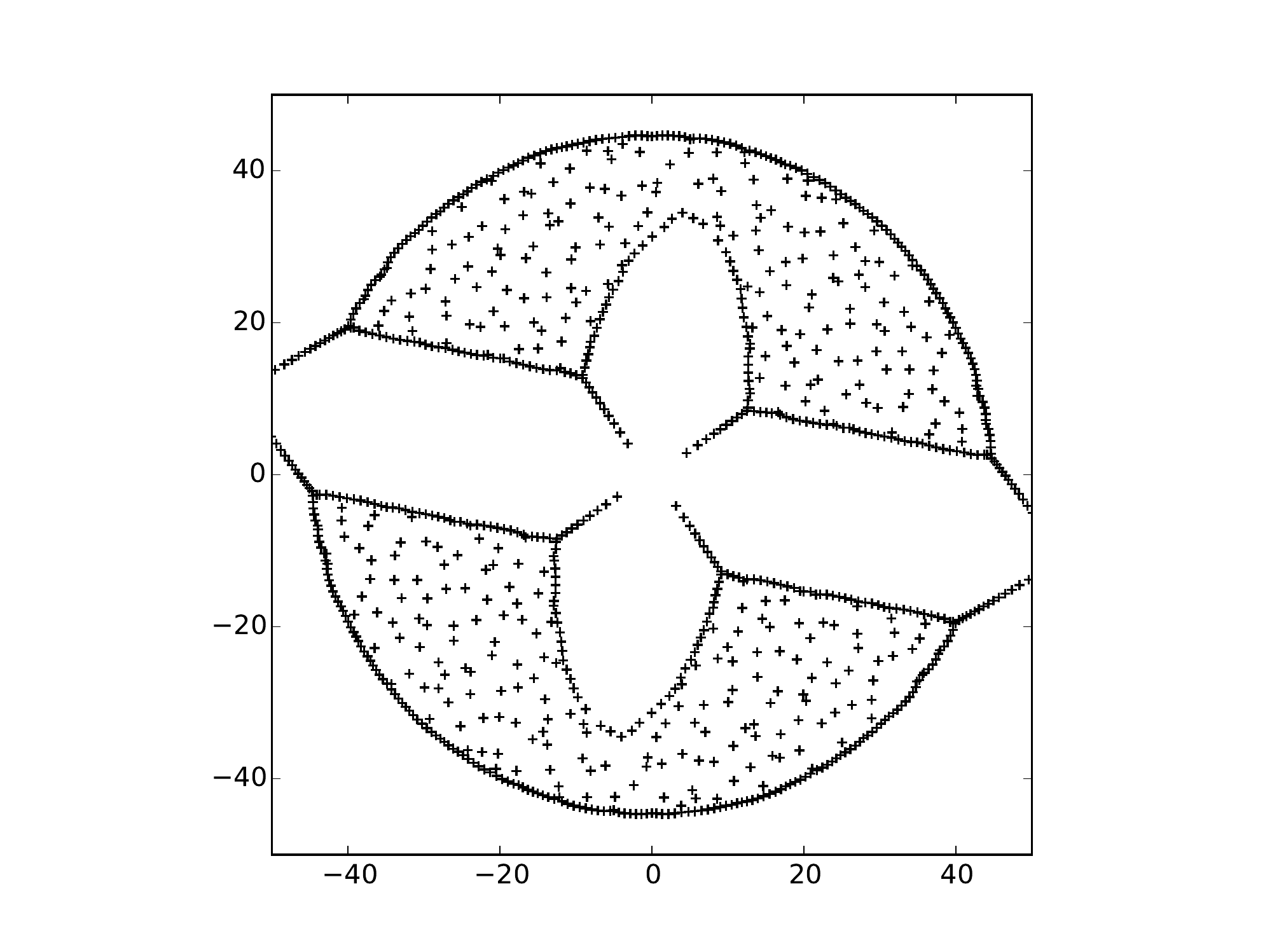}
\textbf{$n=2$}
\includegraphics[trim=0mm 5mm 0mm 10mm,clip,width=\linewidth]{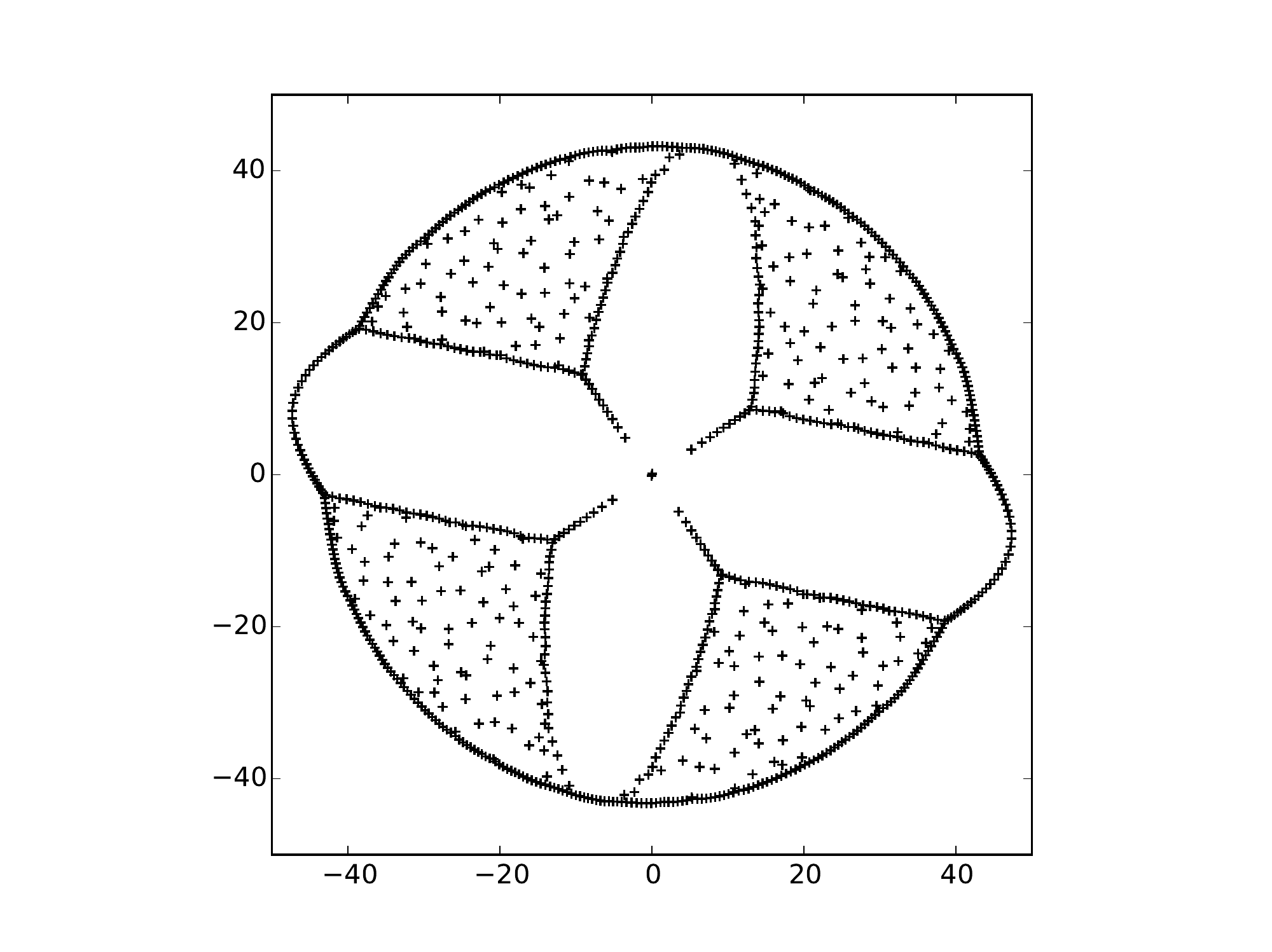}
\textbf{$n=10$}
\includegraphics[trim=0mm 10mm 0mm 10mm,clip,width=\linewidth]{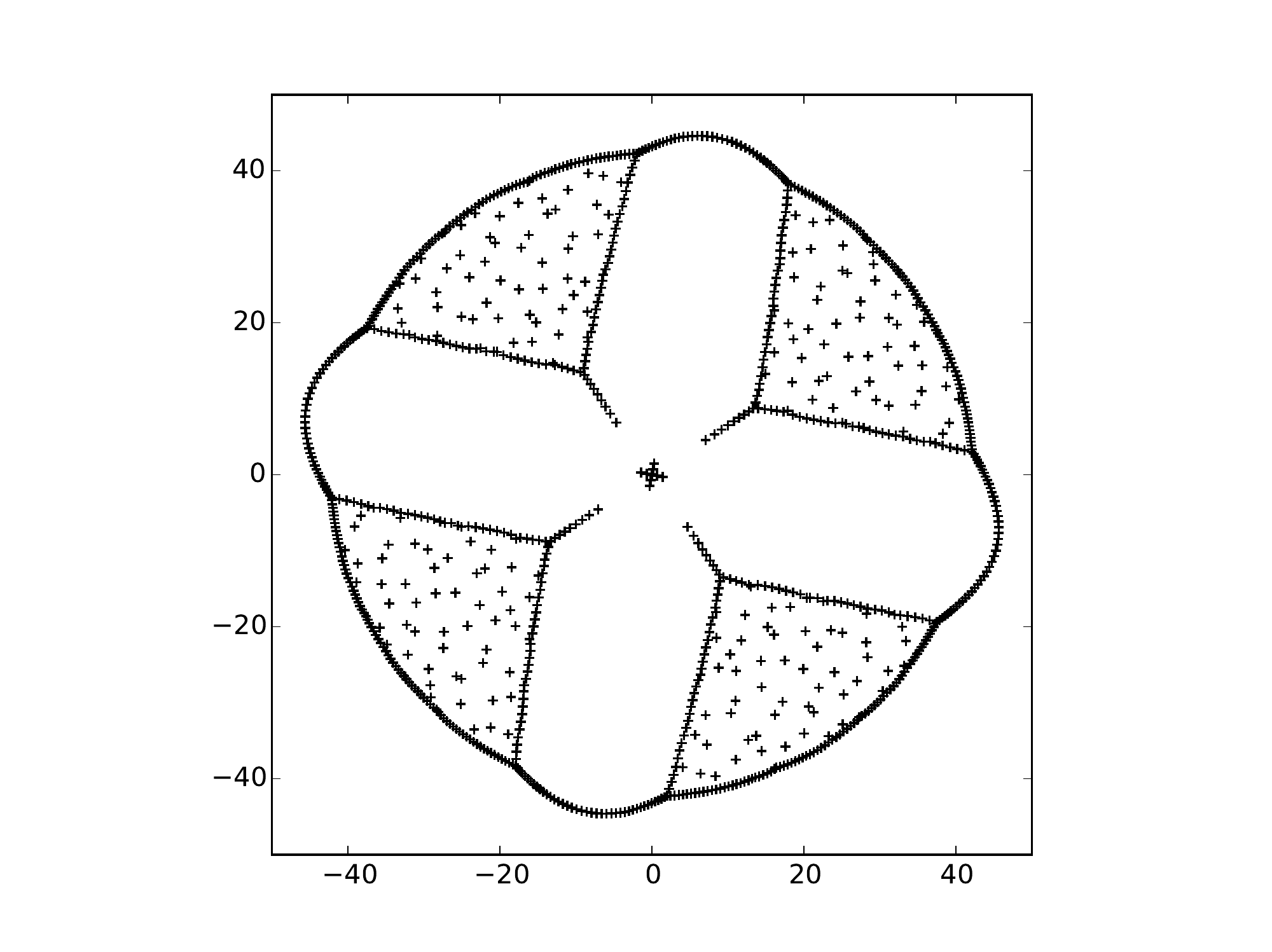}
\caption{Restoration of the $C_4$ symmetry as the Landau level index $n$ increases. Here we adopted a different parametrization of the quartic terms. The Hamiltonian assumes the form $20p_x^2+20p_y^2+2p_x^4+3p_y^4+4\{p_x^2,p_y^2\}+\{p_x,p_y^3\}$.}
\label{fig:crossover}
\end{figure}

\begin{figure}[]
\includegraphics[trim=10mm 0mm 10mm 0mm, clip,width=\linewidth]{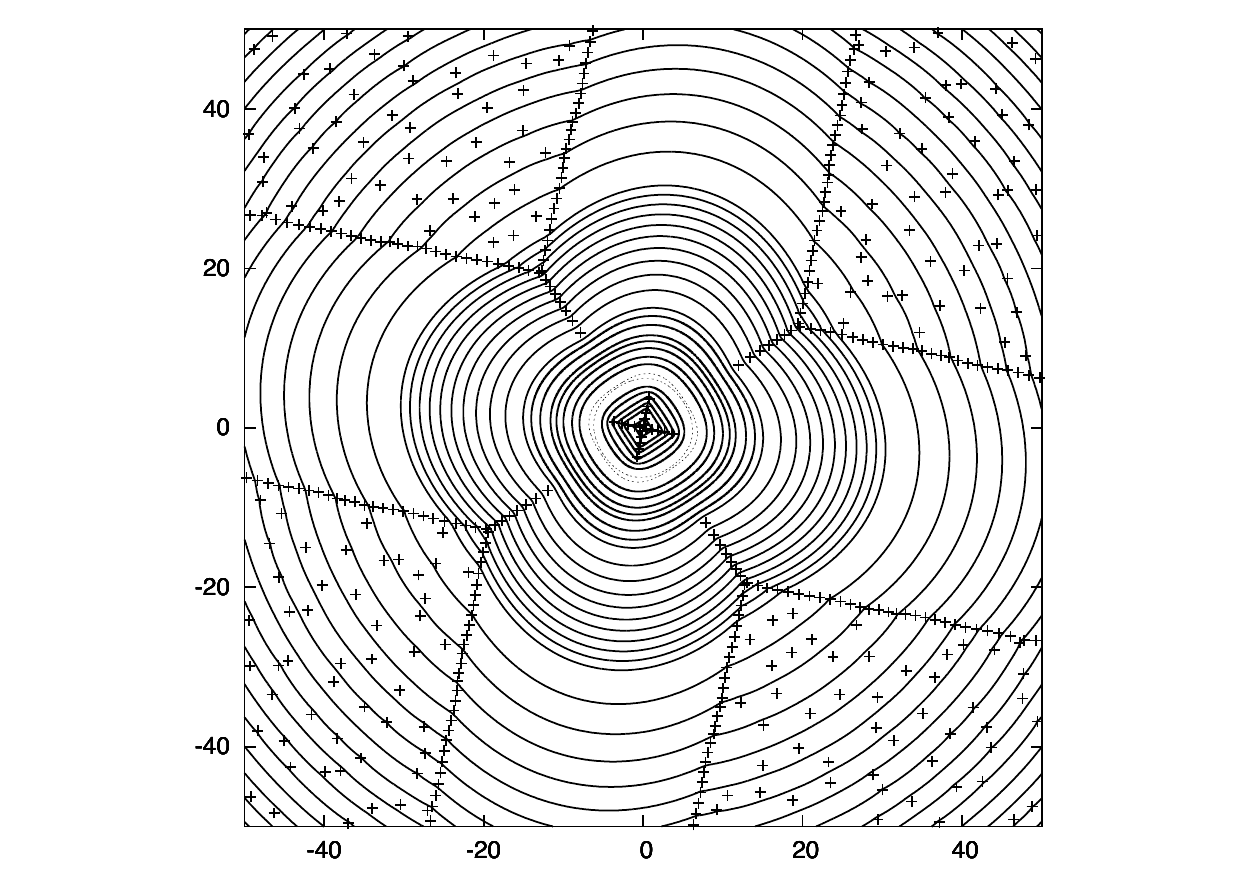}

\bigskip
\includegraphics[trim=10mm 0mm 10mm 0mm, clip,width=\linewidth]{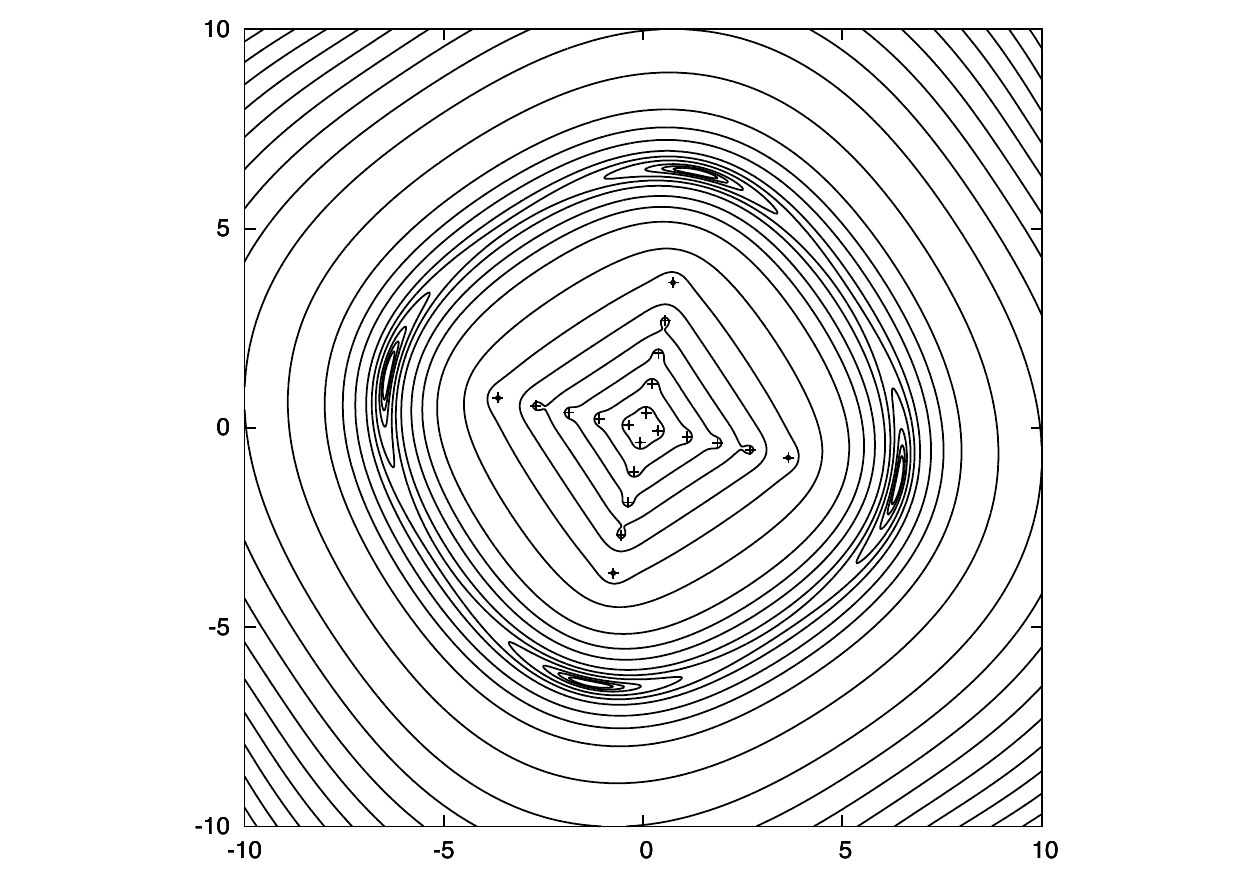}
\caption{Contour plots of $\ln |\Psi_{20}|^2$ for the Hamiltonian $p_x^2+p_y^2+2p_x^4+3p_y^4+4\{p_x^2,p_y^2\}+\{p_x,p_y^3\}$. The lower plot is a zoom-in on the central region. The contours have the same piece-wise structures as in the case of pure quartic terms. Also note that the central cross still points to the maxima, and that the spikes to the saddle points.}
\label{fig:contours}
\end{figure}
We treated the problem numerically by  projecting the Hamiltonian into
the Hilbert space spanned by states where $f_n$ are polynomials of
finite degree $N \gg n$,  and solving a  tridiagonal Hermitian
eigenproblem
to obtain the Taylor series for $f_n$,  and then
used a root finder to obtain approximate eigenstates in the form
\begin{equation}
|\psi_n\rangle  \propto \prod_{i=1}^N(a^{\dagger} - z^*_i)|0\rangle.
\end{equation}
To accurately explore the root structure $\{z^*_i\}$,  we carried out the
diagonalization using MPACK,\cite{MPACK} an adaptation of the linear algebra
routines from LAPACK to use the GMP  library for
arbitrary-precision floating-point calculations.     These calculations
reveal the zeroes of $f_n(z^*)$ in some region $|z| < R_N$, with
truncation-dependent features tied to the boundary $|z| \approx
R_N$, but with a structure for $|z| \ll R_N$ that becomes independent
of $N$  as it,  and consequently $R_N$,  is increased, as illustrated in Fig.\ref{fig:cutoff}.   We therefore
believe that this calculational method reveals the true structure of
roots of the holomorphic non-polynomial function $f_n(z^*)$ in
a range $|z| < R$ that can be increased at will at the expense of
increasing the floating-point precision of the numerical
diagonalization.

 Given that the exact eigenstate satisfies a
three-term recurrence relation (or a five-term recurrence relation if quadratic terms are
included)
that depends on its eigenvalue, it is
possible that, once the eigenvalue has been accurately determined,
the Taylor series could be extended  beyond the
truncation point using the recurrence relation, but we found that the
eigenvector of the truncated tridiagonal matrix was accurate enough
for finding the pattern of roots, when calculated with high precision.

A typical layout (Class I) of the roots is presented in Fig.\ref{fig:class1zeroes}. The central zeroes, the number of which determines the topological spin, are organized in a cross shape due to the $C_4$ symmetry. There are $n$ mod 4 degenerate zeroes at the origin, and the degeneracy will be lifted once we add quadratic terms to the Hamiltonian. Another noticeable feature is the four ``spikes" that point to the origin and alternate with the central cross. The regions with quasi-uniform 2D distributions are bounded by line ``charges" with constant density. If the patterns can be extrapolated to infinity, the total number of roots inside $B_r$ will go as $r^2$, in agreement with the mathematical bound mentioned above. The precise locations of the zeroes inside those 2D regions are sensitive to perturbation of the Hamiltonian, but we believe the very existence of such ``dark" areas is of universal significance.

\begin{figure}[]
\includegraphics[trim=100mm 5mm 100mm 15mm,clip,width=\linewidth]{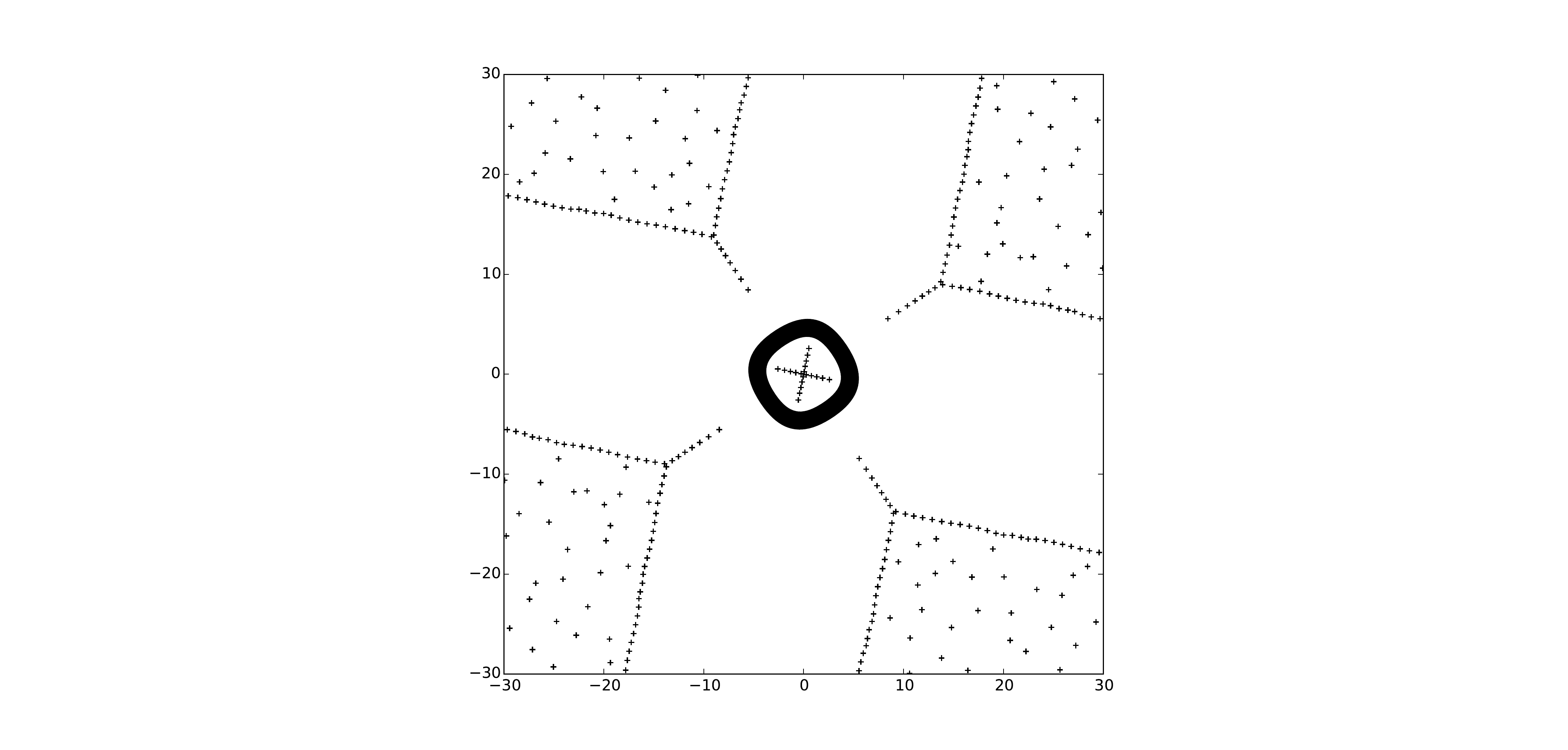}
\caption{Black region with 90 percent of the total weight of the wavefunction. The annulus is bounded by contours at the same value of the amplitude and cleanly separates the central zeroes from the rest of the structure.}
\label{fig:90percent}
\end{figure} 
The structure of the roots strongly suggests that  there is a semiclassical treatment in which $\ln f_n(z)$ is approximated by a  piecewise-holomorphic function in regions separated by branch cuts along the apparent lines of zeroes, with a modification of the background 2D charge-density
distribution of the Poisson equation in the regions where the roots
seem to be  a quasi-uniform 2D distribution. To visualize that, we show in Fig.\ref{fig:class1contours} the contours of the 2D Coulomb potential $V(z,z^*)$, which is $\ln|f_n(z)|^2$ plus a quadratic potential derived from the Gaussian factor. Note that the actual topography has a ``volcano crater" at the center enclosed by an annulus-like region containing the local maxima and saddle points. The fact that all the central zeroes are inside the crater and separated from the rest of the structure is critical for a semiclassical definition of the topological spin. We will see shortly that the same statement doesn't apply to classes II, III, or IV. Additionally, for class I there is a free parameter $c-1$ that controls the shapes of contours of constant $\varepsilon(\bm p)$ in momentum space. A simple calculation reveals that $c-1>2$ corresponds to contours  with concave shapes, while $c-1<2$ to convex ones.

Now we move on to the other three classes of quartic terms. As can be seen from the definitions, their qualitative difference from class I is that the classical dispersion $\varepsilon(\bm{p})$ is consistently zero along one (for classes II, IV) or two (for class III) radial directions on the momentum plane. As a result, the classical orbits for these three classes are not closed. Though the noncommutativity of $p_a$ gives rise to effective quadratic terms for classes II and III, thus closing the quantum orbits, we expect that a clean separation between the central zeroes and the outer structure by a set of contours \textit{cannot} be achieved for any of the three classes, which is verified in Fig.\ref{fig:classes234}.

To complete the discussion, we add quadratic terms to the Hamiltonian. The parametrization now is
\begin{align}
H=\,&Ap_x^2+Bp_y^2+Cp_x^4+Dp_y^4 \notag \\
&+E\{p_x^2,p_y^2\}+F\{p_x,p_y^3\}+G\{p_x^3,p_y\}.
\end{align}
The $C_4$ symmetry is generically not present. However, as we go to high-energy states, the quartic terms will eventually dominate. Therefore we expect to see a quasi-$C_4$ symmetry gradually appear as the Landau index $n$ increases, as shown in Fig.\ref{fig:crossover}. Due to this asymptotic $C_4$ symmetry, the generic features of the root structure and the contours still hold, as illustrated in Fig \ref{fig:contours}. 

We conclude this section by revisiting the semiclassical interpretation of the root structure and demonstrating that the region we pick to isolate the central zeroes can indeed encompass as much as 90 percent of the total weight of the wavefunction, as shown in Fig.\ref{fig:90percent}. In the rotationally invariant case, $\Psi_n$ has $n$ zeroes at the origin, a rim of maxima at a radius of order $\sqrt{n}$ and a Gaussian tail extending to infinity. With a moderate perturbation that breaks rotational invariance, the new zeroes will only appear where the original wavefunction has vanishingly small amplitudes, i.e., around the origin or along the Gaussian tail. This explains why we are able to extract a region that accounts for as much as 90 percent of the total probability while covering no zeroes.

\section{Other properties of the Landau orbit}
\label{sec:properties}

There are some other important characteristic physical properties
of the Landau orbit.    The simplest is its diamagnetic magnetic
moment normal to the plane, given by
\begin{equation}
\mu_n = \left .\frac{\partial E_n}{\partial B}\right
|_{\varepsilon(\bm p)}
\end{equation}
which in the Newtonian model with Galilean invariance reduces to
 $\mu_n$ = $\hbar eB s_n/m_e$.

The second property is the effective mass that characterizes the kinetic energy
of a guiding center that flows along a line of constant $E_n(\bm x)$,
if there is slow adiabatic spatial variation of the Hamiltonian.
This is given by the linear response to a perturbation that breaks the
inversion symmetry of the Landau orbit around its guiding center.
The perturbed Landau levels are given by
\begin{equation}
(H  - v^ap_a)|\Psi_{n,\alpha}(\bm v)\rangle = E_n(\bm
v)|\Psi_{n,\alpha}(\bm v)\rangle,
\end{equation}
where $\bm v$ is a parameter of the perturbed Hamiltonian.
Then
\begin{equation}
\langle \Psi_{n\alpha}(\bm v)|H|\Psi_{n\alpha}(\bm v)\rangle  = E_n +
{\textstyle\frac{1}{2}}m^n_{ab} v^av^b + O(v^4),
\end{equation}
and
\begin{equation}
E_n(\bm v) = E_n - {\textstyle\frac{1}{2}}m^n_{ab} v^av^b + O(v^4),
\end{equation}
For the Newtonian model, $m^n_{ab}$ = $m_e\delta_{ab}$.    Though both
$m^n_{ab}$ and $g^n_{ab}$ are proportional to the
Euclidean metric in the Newtonian model, there is in general no
proportionality between them unless there is a discrete three-, four-
or six-fold crystal rotational symmetry of the lattice plane on which
the electrons move (and no tangential component of the magnetic flux), 
in which case they both remain proportional to the
Euclidean metric which is invariant under the crystal symmetry.

The generic lack of any  relation between the viscosity tensor and the
effective mass tensor can
be traced to the fact that the viscosity derives from perturbations
that mix  Landau levels with the same parity, while the effective mass
derives from mixing of Landau levels with opposite parity.

Since $p_a$ = $\hbar\epsilon_{ab}(x^a - R^a)/\ell_B^2$,  the
perturbation can be written as
\begin{equation}
H(\bm v) =   H - \epsilon_{ab} \hbar v^a(x^b - R^b)/\ell_B^2.
\end{equation}
The effective mass tensor can therefore be interpreted as the part of
the electric susceptibility associated with polarization of the Landau
orbit (giving  it an electric dipole moment defined as the displacement
of its center of charge relative to the guiding center), with $v^a$ =
$\epsilon^{ab}E_b/B$
(so $E_a$ are the tangential components of the electric field in the comoving
frame).
The Landau-orbit polarizability  contribution to the electric susceptibility per unit area is
\begin{equation}
\chi_E^{ab}(\bm x) = \epsilon^{ac}\epsilon^{bd} \frac{e^2\ell_B^2}{2\pi \hbar^2}
\sum_n \nu_n m^n_{cd}(\bm x).
\end{equation}

\section{density and current response}\label{sec:response}
With the properties of Landau levels introduced in previous sections,
we now present a generic version of the electromagnetic responses of
the particle density and current in the absence of rotational
symmetry.   These results agree with those found earlier\cite{Yang} to
order $q^2$, but
which have remained unpublished.  However, we  felt it was useful to
provide (in Appendix \ref{app:A}) a fuller and more
 explicit derivation than that available in
Ref.\onlinecite{Yang}, using a somewhat different formalism to obtain
the same results, and extending Ref.\onlinecite{Yang} to give the
formal expansion of the current-density operator of a non-Galilean system to all orders in $q$.

Denote the particle density and current in the $n$th Landau level by $J_n^a$ and $J_n^0$. Our main results are
\begin{align}
eJ_n^a &=\frac{e^2}{2\pi \hbar}(1+\sigma^{cd}\partial_c\partial_d)\epsilon^{ab}E_b-\chi_m\epsilon^{ab}\partial_b B,\\
eJ_n^0 &=\frac{e^2B}{2\pi \hbar}(1+\sigma^{ab}\partial_a\partial_b\ln B)-\chi_E^{ab}\partial_a E_b,
\end{align}
where
\begin{align}
\sigma^{ab}&=&&\mkern-28mu\frac{1}{6}s_n g_n^{ab}-\left[\frac{1}{6}(\tilde R^a,\{\tilde R^b,dh\})_n+a\leftrightarrow b\right] \notag \\
&&&\mkern-38mu+\frac{1}{4}(\{\tilde R^a, \tilde R^b\},dh)_n,\\
s_n g_n^{ab}&=&&\mkern-28mu\frac{1}{2}\langle \{ \tilde R^a,\tilde R^b\} \rangle_n,\\
\chi_m&=&&\mkern-28mu\frac{e^2l_B^2}{2\pi \hbar^2}\left(\langle dh+d^2h\rangle_n + (dh,dh)_n\right).
\end{align}
The notations used here are shown as follows:
\begin{align}
(A,B)_{n}
&\equiv \sum_{n'\ne n}
\frac{\langle n|A |n'\rangle \langle n'|B|n\rangle +
\langle n|B |n'\rangle \langle n'|A|n\rangle }
{E_n - E_{n'}},\\
\langle A \rangle_n &\equiv \langle n | A | n\rangle,
\end{align}
where $|n\rangle$ and $E_n$ are the $n$th eigenstate and eigenvalue of the unperturbed single-particle Hamiltonian $h$. The scale derivative is defines as
\begin{align}
dh=B_0\frac{\partial h}{\partial B_0}, \quad
d^2 h= B_0 \frac{\partial (dh)}{\partial B_0}.
\end{align}
The first term in $\sigma^{ab}$ is proportional to the inverse Hall viscosity tensor and therefore universal. The second (third) term is due to Landau-level mixing between states with opposite (the same) parity, thus depending on the details of the Hamiltonian. When there is rotational symmetry, the third term vanishes because the eigenstates are invariant with respect to rescaling of $B_0$. When Galilean symmetry is present with a universal (inverse) metric $g^{ab}$, the second term can be explicitly calculated,
\begin{align}
-[\frac{1}{6}(\tilde R^a,\{\tilde R^b,dh\})_n+a\leftrightarrow b] = \frac{4}{3}s_n g^{ab},
\end{align}
which combined with the first term gives
\begin{align}
\sigma^{ab}=\frac{3}{2}s_n g^{ab}.
\end{align}
In the Euclidean case where $g^{ab}=l_B^2 \delta^{ab}$, it agrees with existing results in the literature.\cite{Abanov, Son, Hoyos, Rudro}

\section{conclusion}
\label{sec:conclusion}
We have studied the geometry of Landau orbits in the absence of
Galilean and rotational symmetries. With the help of consistent index
placement, we reformulated the Hall viscosity and used it as a natural
choice for the unimodular metric $\tilde g^n_{ab}$ that describes the
shape of each Landau orbit. We thus demonstrated that the metric in
the IQHE is derived from the intrinsic dynamics of the system rather
than induced from the 3D space the system is embedded in. By
investigating the root structure of the holomorphic part of the
single-particle wavefunction, we defined a topological spin
$s_n=n+\frac{1}{2}$ where $n$ is the number of central zeroes enclosed
by a zero-free region that can be thought of as a ``fattened"
semiclassical orbit. Compared to the ``intrinsic orbital angular
momentum''interpretation, this definition is more generic and captures the topological nature of this quantity. We also introduced a mass tensor $m^n_{ab}$ associated with the kinetic energy of the guiding centers that only coincides with $\tilde g^n_{ab}$ when there is a continuous rotational symmetry.

With all the ingredients above, we presented the generic (without rotational symmetry) results for particle density and current responses. The results clearly showed that the Hall conductivity at finite wavevector consists of a universal term proportional to the inverse Hall viscosity tensor and terms related to Landau-level mixing. We believe that a generic version of the effective action should also possess these features which are usually hidden by the simplification of applying special symmetries.
\section{acknowledgments}
This work was supported by DOE grant DE-SC0002140 and the W. M. Keck Foundation.

\onecolumngrid
\newpage
\appendix
\section{Density and current reponses in the absence of rotational symmetry}\label{app:A}
Let  the two components of the dynamical momentum $\bm p$ = $\bm
p^{\dagger}$ 
have the
commutation relation
\begin{equation}
[p_a,p_b] = i\hbar^2 \ell_B^{-2}\epsilon_{ab},   
\label{heis}
\end{equation}
where the antisymmetric  Levi-Civita symbol $\epsilon_{ab}$ =
$-\epsilon_{ba}$ has $\det \epsilon$ = 1.
With the Schr\"odinger representation
\begin{equation}
p_a = -i\hbar \partial_a - eA_a(\bm x), \quad \partial_a \equiv
\frac{\partial}{\partial x^a},
\end{equation}
this represents the momentum of a charge-$e$ particle
moving in a uniform background magnetic flux density 
 \begin{equation}
B_0 = 
\partial_1 A_2(\bm x) -  \partial_2 A_1(\bm x) =   
\mathop{\rm Pf}(\epsilon) (\hbar/e)\ell_B^{-2},
\end{equation}
where  the Pfaffian $\mathop{\rm Pf}(\epsilon)$ = $\epsilon_{12}$
of the Levi-Civita symbol
is odd under time-reversal.    
A metric-independent formalism that consistently 
distinguishes between covariant (lower) and contravariant (upper)
2D spatial indices will be used here.
The dual symbol $\epsilon^{ab}$ is defined by
\begin{equation}
\epsilon^{ab}\epsilon_{cd} = \delta^a_c\delta^b_d
-  \delta^a_d\delta^b_c,
\end{equation}
where $\delta^a_b$ is the Kronecker symbol.
In a metric-independent formalism, there is no symbol ``$\delta_{ab}$'',
which if encountered,  would represent the Euclidean metric in  a Cartesian
coordinate system.

The algebra (\ref{heis}), in which the commutator is a c-number, 
is the Heisenberg algebra $\mathfrak h_2$.   Its universal
enveloping algebra $ \mathbb U(\mathfrak h_2)$ is spanned by the basis
\begin{equation}
\{ 1, p_a, \{p_a,p_b\}, \{p_a,p_b,p_c\} , \dots\}
\end{equation}
where $\{A,B,\ldots, C\}$ is the symmetrized product of $n$  operators
normalized so that $\{\lambda A,\lambda A,\ldots,\lambda A\}$ =
$n!(\lambda A)^n$.
The kinetic energy $h$ can then be given in the form
\begin{equation}
h = H(\bm p)  \equiv  \sum_n \frac{1}{n!} H_n^{a_1,a_2,\ldots a_n}p_{a_1}p_{a_2}\ldots p_{a_n}
\end{equation}
where the coefficients $H_n^{a_1\ldots a_n}$ are fully symmetric in
all indices.

It is useful to define
\begin{equation}
\tilde R^a  = (\tilde R^a)^{\dagger}= -\hbar^{-1}\ell_B^2 \epsilon^{ab}p_b.
\end{equation}
Then
\begin{eqnarray}
{[}R^a,p_b] &=& i\hbar \delta^a_b, \\
{[}\tilde R^a,\tilde R^b] &=& i\ell_b^2 \epsilon^{ab}.
\end{eqnarray}
The unitary boost operator, parametrized by a real c-number vector $\bm q$,
is given by
\begin{equation}
U(\bm q) = U(-\bm q)^{\dagger} = e^{i\bm q\cdot \tilde {\bm R}}, \quad [q_a,q_b] = 0.
\end{equation}
Note that the scalar product $\bm q\cdot \tilde {\bm R}$ $\equiv$  $q_a\tilde
R^a$ can only be formed between a covariant and a contravariant vector.
Then
\begin{equation}
U(-\bm q)\bm pU (\bm q) = \bm p + \hbar\bm  q.
\end{equation}
and the derivatives $h^{a_1a_2\ldots a_n}$ of $h$
are defined by
\begin{equation}
h(\bm q) \equiv   U(-\bm q)hU(\bm q) = \sum_{n=0}^{\infty}
\frac{\hbar^n}{n!}h^{a_1a_2\ldots a_n}q_{a_1}q_{a_2}\ldots q_{a_n},
\end{equation}
where the tensor-valued operators $h^{a_1a_2\ldots a_n}$ are fully
symmetric in their indices.
In particular, the first derivative $h^a$ is the velocity
operator
$v^a$.
Note also that
\begin{equation}
U(-\bm q)U(-\bm q') \bm p U(\bm q') U(\bm q) = U(-(\bm q + \bm q')) \bm p
U(\bm q + \bm q').
\end{equation}
From this
\begin{equation}
U(- \bm q)h^{a_1,\ldots a_m}U(\bm q) =
\sum_{n=0}^{\infty}\frac{\hbar^n}{n!} h^{a_1\ldots a_m b_1\ldots
  b_n}q_{b_1}\ldots q_{b_n}.
\end{equation}
as in the differentiation of c-number functions, where $(f')'$ = $f''$.

Now let $\bm a$ $\in$ $\mathbb U(\mathfrak h)$ be a  vector-valued operator  and
$\lambda$ be a c-number.
Then define the expansion
\begin{equation}
H(\bm p +\lambda \bm a) = h+ \lambda h_1(\bm a) +
O(\lambda^2).
\end{equation}
The quantity $h_1(\bm a)$ can be expressed in two equivalent
ways as the nested-commutator expansions
\begin{eqnarray}
\label{eqn:expansion}
h_1(\bm a) &=&
a_{a_1}h^{a_1} + \frac{1}{2!}[p_{a_1}a_{a_2}]h^{a_1a_2}
+ \frac{1}{3!} [p_{a_1},[p_{a_2},a_{a_3}]] h^{a_1a_2a_3}+
                      \ldots\\
&=& h^{a_1}a_{a_1}
+ \frac{1}{2!} h^{a_1a_2}[a_{a_1},p_{a_2}]
+ \frac{1}{3!} h^{a_1a_2a_3} [[a_{a_1},p_{a_2}],p_{a_3}] +
    \ldots .
\end{eqnarray}

The operator $h$ is also a function of the scale factor $B_0$,
and more properly should be written as $h(B_0)$.  The
magnetization operator $m(\bm p)$ is defined by
\begin{eqnarray}
m &\equiv&  -\left .\frac{\partial h}{\partial B_0}\right |_{\{H_n\}}
= -{\textstyle\frac{1}{2}}B_0^{-1}h_1(\bm p) \nonumber \\
&=& -{\textstyle \frac{1}{2}}B_0^{-1} \bm v\cdot \bm p =
-{\textstyle \frac{1}{2}}B_0^{-1} \bm p\cdot \bm v
={\textstyle\frac{1}{2}} B_0^{-1}\left (h + i\ell_B^{-2}\epsilon_{ab}\tilde R^ah\tilde
    R^b\right ).
\end{eqnarray}
If the kinetic energy is a quadratic operator, $h$ =  $\frac{1}{2}H_2^{ab}p_ap_b$,
then $B_0m$ = $-h$. For any scalar operator $h$ $\in$ $\mathbb U(\mathfrak h_2)$, one can
define the scale 
derivative
\begin{equation}
dh \equiv B_0 \frac{\partial h}{\partial B_0} = {\textstyle\frac{1}{2}}
\left ( -i \ell_B^{-2}\epsilon_{ab}\tilde R^ah\tilde R^b - h\right ),
\end{equation}
so
\begin{equation}
dh = - B_0m.  
\end{equation}
Higher orders of derivatives can be recursively defined as
\begin{equation}
d^{n+1}h\equiv B_0\frac{\partial(d^{n}h)}{\partial B_0}, \quad d^1h\equiv dh.
\end{equation}
Note that if $Q$ $\equiv$ $\bm q\cdot \tilde {\bm R}$ $\in \mathfrak h$, then
\begin{equation}
Qh \tilde R^a - \tilde R^ahQ = [Q,R^a](h+2dh)
\label{mag}
\end{equation}
where $[Q,R^a]$ is a c-number.

Now enlarge the algebra to add a second independent Heisenberg
algebra, that of the \textit{guiding centers}:
\begin{equation}
[R^a,R^b] = -i\epsilon^{ab}\ell_B^2, \quad [R^a,\tilde R^b] = 0.
\end{equation}
then
\begin{equation}
\bm r = \bm R + \tilde {\bm R}
\end{equation}
is the classical coordinate of the charged particle:
\begin{equation}
[r^a,r^b] = 0, \quad [r^a,p_b] = i\hbar \delta^a_b.
\end{equation}
Note that
\begin{equation}
\mathop{\rm Tr}(e^{i\bm q\cdot \bm R}) = 2\pi \delta^2(\bm q\ell_B)
\end{equation}
where the trace is over the Hilbert space of the guiding-center, and
$\delta^2(\bm x)$ is the 2D Dirac delta-function.

Now introduce an additional vector potential
\begin{equation}
A_a(\bm x)  =\int \frac{d^2\bm q}{(2\pi)^2} \tilde A_a(\bm q) e^{i\bm
  q\cdot \bm x},
\end{equation}
with 
\begin{equation}
\delta B(\bm x) =  \partial_1A_2(\bm x) - \partial_2A_1(\bm x).
\end{equation}
Now define the operator
\begin{equation}
A_a \equiv A_a(\bm r) = \int \frac{d^2\bm q}{(2\pi)^2} \tilde A_a(\bm q) e^{i\bm q\cdot
  \bm R} U(\bm q).
\end{equation}
The problem of the charged  particle moving in the 2D plane $\bm x$ in
the presence of a non-uniform magnetic flux density $B_0 + \delta
B(x)$ is then described by the Hamiltonian
\begin{equation}
H = H(\bm p  -e\bm A)
\end{equation}
The electric charge density operator is given by
\begin{equation}
J^0(\bm x) = e\int \frac{d^2\bm q}{(2\pi)^2} e^{i\bm q(\bm R-\bm x)}
U(\bm q),
\end{equation}
and  the electric current-density
operator is given by
\begin{equation}
J^a(\bm x) =  e\int\frac{d^2\bm q}{(2\pi)^2} e^{i\bm q\cdot (\bm R-\bm
  x)} J^a(\bm q),
\end{equation}
where, to leading order in $\tilde {\bm A}$,
\begin{equation}
e_a J^a(\bm q) = 
h_1(\bm eU(\bm q)) + O(\tilde {\bm A}),
\end{equation}
where $\bm e$ is a c-number unit vector.
 
The continuity equation (gauge invariance) requires that
\begin{equation}
(i\hbar)^{-1}[J^0(\bm x),H] + \partial_aJ^a(\bm x) = 0.
\end{equation}
This must be satisfied at each order of an expansion in $\tilde
A_a(\bm q)$.  In the limit of uniform magnetic flux density $B_0$, 
the continuity relation is
\begin{equation}
[h,U(\bm q)] = \hbar q_aJ^a(\bm q),
\end{equation} 
and the functional form of the expansion of $J^a(\bm q)$ must
be
\begin{equation}
\hbar J^a(\bm q) =  -i [\alpha^a(\bm q),h] + 
i\epsilon^{ba}q_b\ell_B^2\beta(\bm q).
\end{equation}
where
\begin{equation}
iq_a\alpha^a(\bm q) =   U(\bm q) - 1
\end{equation} 
or
\begin{equation}
\alpha^a(\bm q) = -i \frac{\partial}{\partial q_a}\left (\sum_{n=1}^{\infty}
\frac{1}{n} \frac{(i\bm q\cdot \tilde {\bm R})^n}{n!}\right )
= \sum_{n=1}^{\infty} \frac{1}{n}\frac{1}{n!}\sum_{k=1}^n (iQ)^{k-1}
  \tilde R^a(iQ)^{n-k}.
\label{univ}
\end{equation}

Using (\ref{eqn:expansion}) the explicit form of $J^a(\bm q)$ can be written as 
\begin{align}
i\hbar J^a = U(\bm q)\left( [\tilde R^a,h] +\frac{1}{2!}[-iQ,[\tilde R^a,h]] + \frac{1}{3!}[-iQ,[-iQ,[\tilde R^a,h]]] + \cdots \right),
\end{align}
where $Q=\bm q \cdot \tilde{ \bm R}$ as mentioned above.
Expanding $U(\bm q)$ and the nested commutators gives
\begin{align}
i\hbar J^a = \sum_{m=0}^{\infty} \frac{(iQ)^m}{m!} \sum_{n=0}^{\infty} (-1)^n \frac{1}{(n+1)!} \sum_{k=0}^{n} (-1)^k \binom nk (iQ)^{n-k} [\tilde R^a,h] (iQ)^k.
\end{align}
The coefficient $C_{l,k}$ for the term
\begin{align}
(iQ)^l[\tilde R^a,h](iQ)^k \notag
\end{align}
in the expansion is (using $l+k=m+n$)
\begin{align}
C_{l,k}&=\sum_{m=0}^{l}\frac{1}{m!} (-1)^{l+k-m} \frac{1}{(l+k-m+1)!} (-1)^k \binom{l+k-m}{k} \notag\\
&=\frac{1}{k!l!}\sum_{m=0}^{l} (-1)^{m} \frac{1}{k+m+1}\binom {l}{m}\notag\\
&=\frac{1}{(k+l+1)!}.
\end{align}

Now a significantly simplified form for $J^a(\bm q)$ can be written down
\begin{align}
i\hbar J^a  = &\sum_{n=0}^{\infty}\frac{1}{(n+1)!} \sum_{k=0}^{n} (iQ)^{n-k}[\tilde R^a,h](iQ)^{k},
\end{align}
which is further transformed into the desired form:
\begin{align}
i\hbar J^a =& \sum_{n=0}^{\infty}\frac{1}{(n+1)!} \left((iQ)^n\tilde R^a h-h\tilde R^a (iQ)^n-i\sum_{k=0}^{n-1} (iQ)^{n-1-k}(Qh\tilde R^a-\tilde R^a hQ)(iQ)^k\right) \notag \\
=& \sum_{n=0}^{\infty}\frac{1}{(n+1)!}\left(\frac{1}{2}\{(iQ)^n,\tilde R^a\}h+\frac{1}{2}[(iQ)^n,\tilde R^a]h-\frac{1}{2}h\{\tilde R^a,(iQ)^n\}+\frac{1}{2}h[(iQ)^n,\tilde R^a] \right. \notag \\ &\left.-i[Q,\tilde R^a]\sum_{k=0}^{n-1} (iQ)^{n-1-k}(h+2dh)(iQ)^k \right) \notag \\
=& \sum_{n=0}^{\infty}\frac{1}{(n+1)!}\left(\left[\frac{1}{2}\{(iQ)^n,\tilde R^a\},h\right]-i[Q,\tilde R^a]\sum_{k=0}^{n-1} (iQ)^{n-1-k}(2dh)(iQ)^k\right.\notag \\ &\left.+i[Q,\tilde R^a]\left(\frac{n}{2}(iQ)^{n-1}h+\frac{n}{2}h(iQ)^{n-1}-\sum_{k=0}^{n-1} (iQ)^{n-1-k}h(iQ)^k\right) \right) \notag \\
=&\, [\alpha^a(\bm q),h]+i[Q,\tilde R^a]\beta(\bm q)
\end{align}
where
\begin{align}
\alpha^a(\bm q)&=\sum_{n=0}^{\infty}\frac{1}{2(n+1)!}\{(iQ)^n,\tilde R^a\},\\
\beta(\bm q)&=-\sum_{n=0}^{\infty}\frac{2}{(n+1)!}\sum_{k=0}^{n-1} (iQ)^{n-1-k}(dh)(iQ)^k+\sum_{n=0}^{\infty}\frac{1}{2(n+1)!}\sum_{k=0}^{n-1} [(iQ)^{n-1-k},[(iQ)^k,h]]
\end{align}

The expansion for $\alpha^a(\bm q)$ agrees with the universal form
(\ref{univ}),
and the leading term in $\beta(\bm q)$, $-dh=B_0m$, is just the  expected magnetization
term in the total current density
\begin{equation}
J^0 = J^0_{\text{free}} - \partial_aP^a, \quad J^a = J^a_{\text{free}}  + \epsilon^{ab}\partial_bM  + \partial_tP^a.
\end{equation}

We are now in a position to give the 
perturbation result for a non-uniform magnetic flux density and a non-zero electric field.
The perturbed Hamiltonian is
\begin{align}
H = &\;h -e\int \frac{d^2\bm q}{(2\pi)^2} \tilde A_0(\bm q)  U(\bm q)e^{i\bm q
  \cdot \bm R} -e\int \frac{d^2\bm q}{(2\pi)^2} \tilde A_a(\bm q)  J^a(\bm q)e^{i\bm q
  \cdot \bm R} \notag \\
&-e^2\int \frac{d^2\bm q}{(2\pi)^2} \tilde A_a(-\bm q) \Gamma^{ab}(\bm q)  \tilde A_b(\bm q)e^{i\bm q
  \cdot \bm R}+\mathcal O(A^3),
\end{align}
where $\Gamma^{ab}(\bm q)$ is a gauge counter-term that is important for ensuring gauge invariance. Its explicit form is:
\begin{align}
\Gamma^{ab}(\bm q)&=\frac{1}{\hbar^2}\sum_{n=0}^{\infty}\frac{1}{(n+2)!}[\overbrace{iQ,\cdots[iQ}^\text{n},[\tilde R^a,[h,\tilde R^b]]]\cdots]+\frac{1}{\hbar^2} \sum_{n=0}^{\infty}\frac{1}{(n+2)!}[\overbrace{-iQ,\cdots[-iQ}^\text{n},[\tilde R^a,[h,\tilde R^b]]]\cdots]\\
&=\frac{1}{\hbar^2}\sum_{k=0}^{\infty}\frac{2}{(2k+2)!}[\overbrace{iQ,\cdots[iQ}^\text{n},[\tilde R^a,[h,\tilde R^b]]]\cdots].
\end{align}
The  perturbed  particle density and current density of a filled Landau level 
is
\begin{align}
J^0_n(\bm x) &=  \frac{1}{2\pi \ell_B^2} \left ( 1- e\int \frac{d^2\bm
    q}{(2\pi)^2}
\chi_n^{00}(\bm q)  \tilde 	U(\bm q) e^{i\bm q \cdot \bm x} - e\int \frac{d^2\bm
    q}{(2\pi)^2}
\chi_n^{0a}(\bm q)  \tilde A_q(\bm q) e^{i\bm q \cdot \bm x}\right ),\\
J^a_n(\bm x) &=  -\frac{1}{2\pi \ell_B^2} \left( e\int \frac{d^2\bm
    q}{(2\pi)^2}
\chi_n^{0a}(-\bm q)  \tilde A_0(\bm q) e^{i\bm q \cdot \bm x}+e\int \frac{d^2\bm
    q}{(2\pi)^2}
\tilde \chi_n^{ab}(\bm q)  \tilde A_b(\bm q) e^{i\bm q \cdot \bm x}\right),
\end{align}
where
\begin{align}
\chi^{00}_n(\bm q) &=
{\sum_{n'(\ne n)}}
\frac{
\langle n| U(-\bm q)|n'\rangle \langle n'|
U^a(\bm q)|n\rangle   
+ \langle n |U(\bm q)
|n'\rangle
\langle n'|U(-\bm q) |n\rangle 
}{E_n-E_{n'}},\\
\chi^{0a}_n(\bm q) &=
{\sum_{n'(\ne n)}}
\frac{
\langle n| U(-\bm q)|n'\rangle \langle n'|
J^a(\bm q)|n\rangle   
+ \langle n |J^a(\bm q)
|n'\rangle
\langle n'|U(-\bm q) |n\rangle 
}{E_n-E_{n'}},\\
\tilde \chi_n^{ab}(\bm q) &= \chi_n^{ab}(\bm q) + \langle n|\Gamma^{ab}(\bm q)|n\rangle ,\\
\chi^{ab}_n(\bm q) &=
{\sum_{n'(\ne n)}}
\frac{
\langle n| J^a(-\bm q)|n'\rangle \langle n'|
J^b(\bm q)|n\rangle   
+ \langle n |J^b(\bm q)
|n'\rangle
\langle n'|J^a(-\bm q) |n\rangle 
}{E_n-E_{n'}}.
\end{align}
Below we will simplify our notation by defining
\begin{equation}
(A,B)_{n}
\equiv \sum_{n'\ne n}
\frac{\langle n|A |n'\rangle \langle n'|B|n\rangle +
\langle n|B |n'\rangle \langle n'|A|n\rangle }
{E_n - E_{n'}},
\end{equation}
so
\begin{align}
\chi_n^{00}(\bm q)&=(U(-\bm q), U(\bm q))_n, \\
\chi_n^{0a}(\bm q)&=(U(-\bm q), J^a(\bm q))_n, \\
\chi_n^{ab}(\bm q)&=(J^a(-\bm q), J^b(\bm q))_n.
\end{align}
We will also denote the expectation value of some operator $\mathcal O$ in the $n$th Landau level by $\langle \mathcal O \rangle_n$.

Using the expansions of $U(\bm q)$ and $J^a(\bm q)$, these correlation functions are readily obtained:
\begin{align}
\chi^{00}_n(\bm q)&=-(Q,Q)_n+\mathcal O(|\bm q|^4)\\
\chi^{0a}_n(\bm q)&= -\frac{1}{\hbar}[Q,\tilde R^a]\left(1-\frac{1}{6}\langle Q^2 \rangle_n +\frac{1}{3}(Q,\{Q,dh\})_n-\frac{1}{2}(Q^2,dh)_n\right)+\mathcal{O}(|\bm q|^5),\\
\tilde \chi_n^{ab}(\bm q)&= -\frac{1}{\hbar^2}[Q,\tilde R^a][Q,\tilde R^b]\left(\langle dh+d^2h\rangle_n + (dh,dh)_n\right)+\mathcal O(|\bm q|^4)
\end{align}
If we define
\begin{align}
\chi^{ab}_E&=-\frac{e^2}{2\pi l_B^2}(\tilde R^a, \tilde R^b)_n,\\
\sigma^{ab}&=\frac{1}{6}s_n g_n^{ab}-\left[\frac{1}{6}(\tilde R^a,\{\tilde R^b,dh\})_n+a\leftrightarrow b\right]+\frac{1}{4}(\{\tilde R^a, \tilde R^b\},dh)_n,\\
s_n g_n^{ab}&=\frac{1}{2}\langle \{ \tilde R^a,\tilde R^b\} \rangle_n,\\
\chi_m&=\frac{e^2l_B^2}{2\pi \hbar^2}\left(\langle dh+d^2h\rangle_n + (dh,dh)_n\right),
\end{align}
then
\begin{align}
\chi^{00}_n(\bm q)&=-\frac{2\pi l_B^2}{e^2}q_aq_b \chi^{ab}_E+\mathcal O(|\bm q|^4),\\
\chi^{0a}_n(\bm q)&= -\frac{1}{\hbar}[Q,\tilde R^a]\left(1-\sigma^{ab}q_aq_b\right)+\mathcal{O}(|\bm q|^5),\\
\tilde \chi_n^{ab}(\bm q)&= -\frac{2\pi}{e^2l_B^2}[Q,\tilde R^a][Q,\tilde R^b]\chi_m+\mathcal O(|\bm q|^4).
\end{align}
Note that $\chi_E^{ab}$ and $g^{ab}$ are just the electric susceptibility tensor and the inverse of the Landau-orbit metric (up to normalization) defined in the main body of the paper. $\sigma^{ab}$ is the tensor for the quadratic term in the $\bm q$ expansion of the Hall conductivity and $\chi_m$ is the magnetic Hall conductivity.

Now using
\begin{align}
[Q,\tilde R^a]&=il_B^2\epsilon^{ba}q_b,\\
E(\bm q)&=iq\tilde A_0(\bm q),\\
\delta B(\bm q)&=i\epsilon^{ab} q_a \tilde A_b(\bm q),
\end{align}
we are able to write down the generic responses (in the absence of rotational symmetry) of charge density and current density against a non-vanishing electric field and a non-uniform magnetic field:
\begin{align}
eJ_n^a &=\frac{e^2}{2\pi \hbar}(1+\sigma^{cd}\partial_c\partial_d)\epsilon^{ab}E_b-\chi_m\epsilon^{ab}\partial_b B,\\
eJ_n^0 &=\frac{e^2B}{2\pi \hbar}(1+\sigma^{ab}\partial_a\partial_b\ln B)-\chi_E^{ab}\partial_a E_b.
\end{align}
The Hall conductivity at finite wavevector $\bm q$ is thus
\begin{align}
\sigma_H^{ab}=\frac{e^2}{2\pi \hbar}(1-\sigma^{cd}q_cq_d)\epsilon^{ab},
\end{align}
and we present here again the generic formula for $\sigma^{ab}$:
\begin{align}
\sigma^{ab}=\frac{1}{6}s_n g^{ab}-\left[\frac{1}{6}(\tilde R^a,\{\tilde R^b,dh\})_n+a\leftrightarrow b\right]+\frac{1}{4}(\{\tilde R^a, \tilde R^b\},dh)_n.
\end{align}
The first term in $\sigma^{ab}$ is proportional to the inverse of the Hall viscosity tensor $\eta^H_{ab}$  and hence universal. The second and the third terms are due to inter-Landau-level mixing between states with opposite and the same parity respectively, so they depend on the details of the Hamiltonian. When there is rotational symmetry, the third term vanishes because the eigenstates are invariant with respect to rescaling of $B_0$. When Galilean symmetry is present, the second term can be explicitly calculated to be
\begin{align}
-[\frac{1}{6}(\tilde R^a,\{\tilde R^b,dh\})_n+a\leftrightarrow b] = \frac{4}{3}s_n g^{ab},
\end{align}
which combined with the first term gives
\begin{align}
\sigma^{ab}=\frac{3}{2}s_n g^{ab}.
\end{align}
The case where $g^{ab}=l_B^2 \delta^{ab}$ agrees with existing results in the literature.


\begin{thebibliography} {10}
\bibitem{laughlin}
R. B. Laughlin, Phys. Rev. Lett. {\bf 50}, 1395 (1983).

\bibitem{Moore}
G. Moore and N. Read,  Nuclear Physics B {\bf 360}, 362 (1991).

\bibitem{Bradlyn1}
B. Bradlyn and N. Read, Phys. Rev. B {\bf 91}, 125303 (2015).

\bibitem{Bradlyn2}
B. Bradlyn and N. Read, Phys. Rev. B {\bf 91}, 165306 (2015).

\bibitem{Abanov}
A. G. Abanov and A. Gromov, Phys. Rev. B {\bf 90}, 014435 (2014).

\bibitem{Gromov1}
A. Gromov and A. G. Abanov, Phys. Rev. Lett. {\bf 113}, 266802 (2014).

\bibitem{Gromov2}
A. Gromov and A. G. Abanov, Phys. Rev. Lett. {\bf 114}, 016802 (2015).

\bibitem{Can}
T. Can, M. Laskin,  and P. Wiegmann, Phys. Rev. Lett, {\bf 113},
  046803 (2014).

\bibitem{Cho}
G. Y. Cho, Y. You, and E. Fradkin, Phys. Rev. B {\bf 90}, 115139
  (2014).

\bibitem{Gromov}
A. Gromov, G. Y. Cho, Y. You, A. G. Abanov, and E. Fradkin,
Phys. Rev. Lett. {\bf 114}, 016805 (2015).

\bibitem{Son}
D. T. Son, arXiv:1306.0638

\bibitem{Avron}
J. E. Avron, R. Seiler, and P.G. Zograf, Phys. Rev. Lett. {\bf 75},
697 (1995).

\bibitem{Read}
N. Read, Phys. Rev. B {\bf 79}, 045308 (2009).

\bibitem{Bradlyn}
B. Bradlyn, M. Goldstein, and N. Read, Phys. Rev. B {\bf 86}, 245309
(2012).

\bibitem{Nakata}
M. Nakata, The MPACK (MBLAS/LAPACK): a multiple precision arithmetic
version of BLAS and LAPACK, version 0.8.0 (2012), see
http://mlapack.sourceforge.net.

\bibitem{book}
I. Laine, {\sl Nevelanna Theory and Complex Differential Equations}
(Walter de Gruyter \& Co., 1993).

\bibitem{Yang}
B. Yang and F. D. M. Haldane (unpublished); B. Yang, Thesis, Princeton
University 2013, arXiv:1312.2630.

\bibitem{Hoyos}
C. Hoyos and D. T. Son,Phys. Rev. Lett. {\bf 108}, 066805 (2012).

\bibitem{biswas}
R. R. Biswas, arXiv:1311.7149.


\end{thebibliography}
\end{document}